\documentclass[manuscript, acmsmall]{acmart}
\usepackage{enumitem}
\usepackage{multirow}
\usepackage{booktabs}
\usepackage{subcaption} 
\usepackage{algorithm}
\usepackage{algpseudocode}
\usepackage{subcaption}
\usepackage{color}
\usepackage{xcolor}

\AtBeginDocument{%
  }

\setcopyright{acmlicensed}
\copyrightyear{2025}
\acmYear{2025}
\acmDOI{XXXXXXX.XXXXXXX}
\acmBooktitle{Woodstock '18: ACM Symposium on Neural Gaze Detection,
 June 03--05, 2018, Woodstock, NY}
\acmISBN{978-1-4503-XXXX-X/2025/04}

\begin{document}

\title{UNGER: Generative Recommendation with A Unified Code via Semantic and Collaborative Integration}

\author{Longtao Xiao}
\affiliation{%
  \institution{School of Computer Science and Technology, Huazhong University of Science and Technology}
  \country{China}
}
\email{xiaolongtao@hust.edu.cn}

\author{Haozhao	Wang}
\affiliation{%
  \institution{School of Computer Science and Technology, Huazhong University of Science and Technology}
  \country{China}
}
\authornote{Haozhao Wang, Cheng Wang, and Ruixuan Li are corresponding authors.}
\email{hz_wang@hust.edu.cn}

\author{Cheng Wang}
\affiliation{%
  \institution{Huawei Technologies Ltd}
  \country{China}
}
\email{wangcheng250@huawei.com}

\author{Linfei Ji}
\affiliation{%
  \institution{Huazhong University of Science and Technology}
  \country{China}
}
\email{jilinfei@hust.edu.cn}

\author{Yifan Wang}
\affiliation{%
  \institution{Huazhong University of Science and Technology}
  \country{China}
}
\email{d202381481@hust.edu.cn}

\author{Jieming	Zhu}
\affiliation{%
  \institution{Huawei Noah’s Ark Lab}
  \country{China}
}
\email{jiemingzhu@ieee.org}

\author{Zhenhua	Dong}
\affiliation{%
  \institution{Huawei Noah’s Ark Lab}
  \country{China}
}
\email{dongzhenhua@huawei.com}

\author{Rui	Zhang}
\affiliation{%
  \institution{School of Computer Science and Technology, Huazhong University of Science and Technology (www.ruizhang.info)}
  \country{China}
}
\email{rayteam@yeah.net}

\author{Ruixuan	Li}
\affiliation{%
  \institution{School of Computer Science and Technology, Huazhong University of Science and Technology}
  \country{China}
}
\email{rxli@hust.edu.cn}

\renewcommand{\shortauthors}{Xiao et al.}

\begin{abstract}
With the rise of generative paradigms, generative recommendation has garnered increasing attention. The core component is the item \textit{code}, generally derived by quantizing collaborative or semantic representations to serve as candidate items identifiers in the context. However, existing methods typically construct separate codes for each modality, leading to higher computational and storage costs and hindering the integration of their complementary strengths. Considering this limitation, \textit{we seek to integrate two different modalities into a unified code}, fully unleashing the potential of complementary nature among modalities. Nevertheless, the integration remains challenging: the integrated embedding obtained by the common concatenation method would lead to underutilization of collaborative knowledge, thereby resulting in limited effectiveness.

To address this, we propose a novel method, named UNGER, which integrates semantic and collaborative knowledge into a \textbf{\underline{un}}ified code for \textbf{\underline{ge}}nerative \textbf{\underline{r}}ecommendation. Specifically, we propose to adaptively \textit{learn} an integrated embedding through the joint optimization of cross-modality knowledge alignment and next item prediction tasks. Subsequently, to mitigate the information loss caused by the quantization process, we introduce an intra-modality knowledge distillation task, using the integrated embeddings as supervised signals to compensate. Extensive experiments on three widely used benchmarks demonstrate the superiority of our approach compared to existing methods.
\end{abstract}




\maketitle
\section{Introduction}
\begin{figure}[h]
  \centering
  \includegraphics[width=\linewidth]{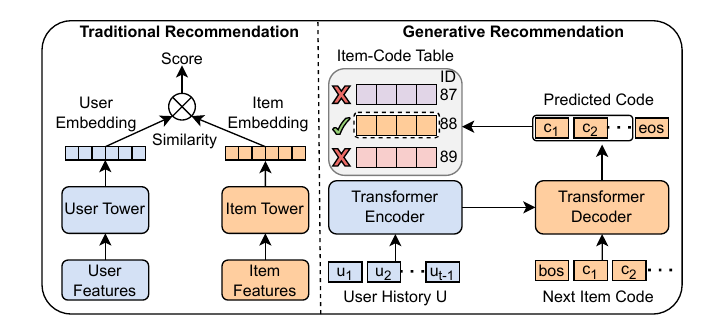}
  \caption{Traditional vs. Generative Recommendation.}
  \label{fig:generative rec}
\end{figure}
To mitigate the issue of information overload, recommendation systems have become widely adopted in modern society, such as in video \cite{zhao2019recommending}, movie \cite{diao2014jointly}, and music \cite{kim2007music} scenarios. Recommendation systems mine users' interests and proactively recommend items that they may be interested in. Modern recommendation systems typically adopt a retrieve-and-rank strategy. During the retrieval process, the model needs to retrieve the most relevant items from a large candidate pool based on the item embeddings learned through representation learning. However, existing approaches, whether based on dual-tower models \cite{covington2016deep, wang2021cross} or graph models \cite{wang2018billion, ying2018graph}, generally rely on dot-product (cosine) similarity with approximate nearest neighbor (ANN) search \cite{andoni2018approximate, arya1998optimal}. Although ANN accelerates top-k recommendation retrieval, its search indices are constructed using tools such as Faiss \cite{johnson2019billion} and SCANN \cite{guo2020accelerating}, which are independent of the optimization process of the recommendation model, thereby limiting the overall effectiveness of recommendation systems \cite{zhu2024cost}. \nocite{DBLP:conf/iclr/WangXLX0024}

\begin{figure}[h]
  \centering
  \includegraphics[width=\linewidth]{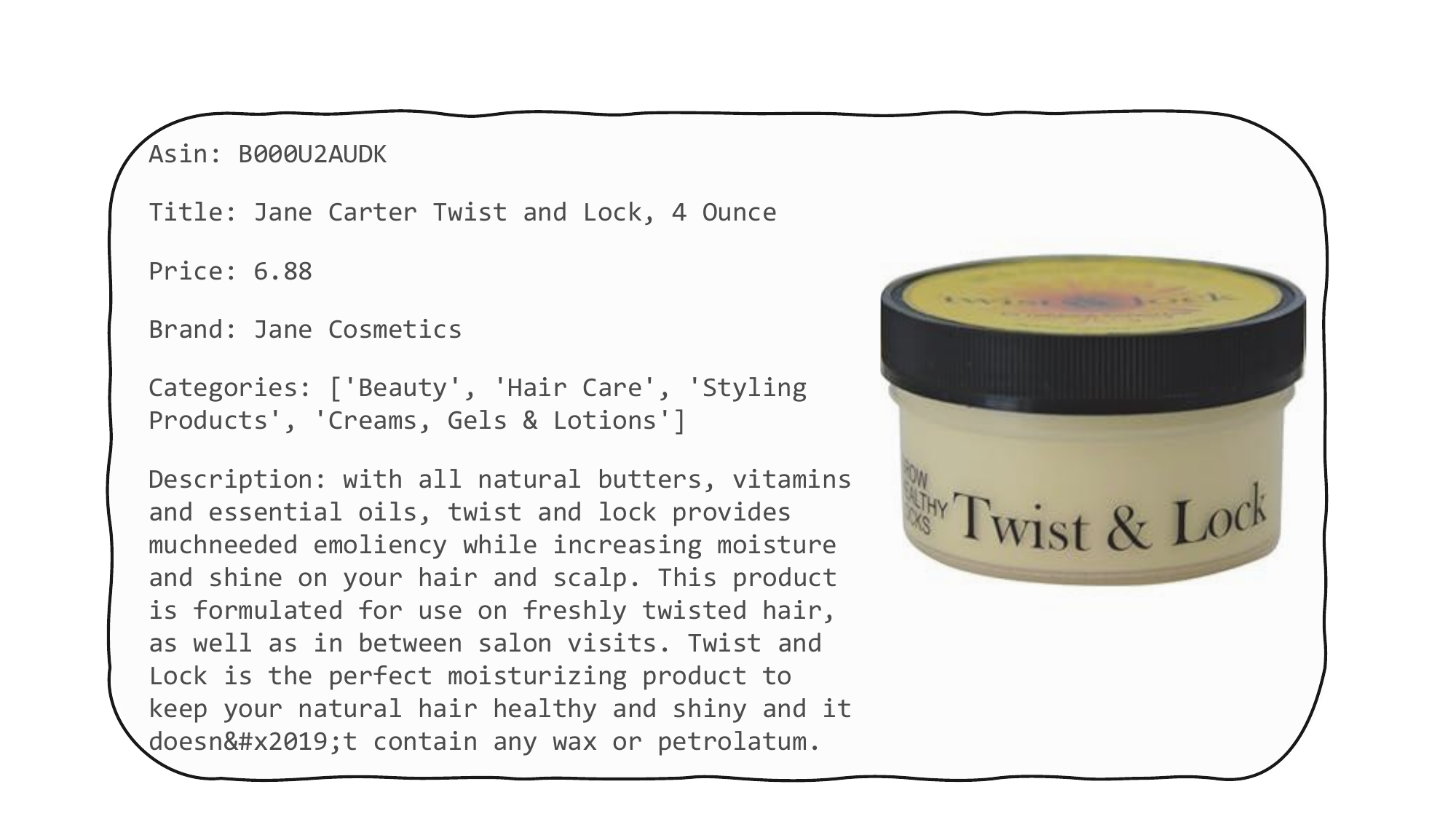}
  \caption{An example of item semantic knowledge.}
  \label{fig:product}
\end{figure}

To address this limitation, generative recommendation \cite{liu2024multimodal, wang2023generative, wang2023diffusion, rajput2023recommender}, which adopts code as the core component and frames the recommendation task as autoregressive code sequence generation, has emerged as a promising research direction owing to its potential to enable more efficient decoding without the need for ANN indexes. Unlike traditional methods that match users and items based on embeddings, generative recommendation predicts candidate item code directly, as depicted in Figure \ref{fig:generative rec}. Specifically, pre-trained semantic encoders \cite{ni2021sentence, lee2018pre} or collaborative encoders \cite{zhou2018deep, kang2018self} are first utilized to encode item semantic or ID information such as shown in Figure \ref{fig:product} into embeddings. Then, embedding quantization \cite{lee2022autoregressive, qi2017effective, li2024embedding} is applied to map these embeddings into discrete codes forming a lookup table of item codes. The input of transformer will be converted into a sequence of codes, enabling the model to autoregressively predict the next item's code step by step. 

Existing methods typically construct independent codes for different modalities knowledge, either collaborative or semantic modality. For instance, Recforest \cite{feng2022recommender} encodes item collaborative knowledge into a single set of codes, whereas TIGER \cite{rajput2023recommender} focuses on encoding item semantic knowledge into a separate code set. Furthermore, EAGER \cite{wang2024eager} encodes both collaborative and semantic knowledge into two independent sets of codes. These approaches have demonstrated that leveraging both collaborative and semantic information is necessary to enhance the effectiveness and robustness of recommendation models. However, for generative recommenders aiming to integrate both semantic and collaborative knowledge, two separate codes are unacceptable, as the dual-code framework significantly increases storage and inference costs, rendering it impractical for large-scale deployments. Besides, the intrinsic misalignment between semantic and collaborative knowledge limits the potential to fully exploit their complementary strengths with two separate codes \cite{kim2024sc}.

To overcome the aforementioned challenges, we propose a novel method, UNGER, which integrates collaborative and semantic knowledge into a unified code for generative recommendation. Unified codes not only reduce storage and computation demands but also accelerate inference and lower latency, which are critical for real-time responses. As illustrated in Figure \ref{fig:speed}, a unified code demonstrates significantly 2.8x faster inference compared to two separate codes. Moreover, dual-code methods inherently treat semantic and collaborative knowledge as independent entities, thereby missing opportunities to harness their synergistic potential. In contrast, a unified code, by integrating both knowledge at an early stage, can more effectively exploit their complementary strengths.

However, it is non-trivial to adopt a unified code. \textit{The core challenge in adopting a unified code lies in how to get a unified representation which integrates knowledge from both collaborative and semantic modalities.} A common practice is to concatenate semantic and collaborative features to create a unified representation. However, this approach suffers from a phenomenon termed \textit{the semantic dominance issue}. As depicted in Figure \ref{fig:concat}, the final representation aligns more closely with the semantic side, leading to underutilization of the collaborative side. This imbalance not only limits the potential of integrating collaborative knowledge but, in extreme cases, can even degrade performance to levels below those achieved with semantic knowledge alone. The root cause of this issue lies in the inability of concatenation to account for the inherent differences between the two different modalities, such as variations in signal strength and representational misalignment \cite{wang2024enhanced}. Consequently, semantic representations dominate the final output, undermining the expected advantages of unified representations.

We analyze the challenges of integrating both collaborative and semantic knowledge into a unified code for generative recommendation and address them from the following three aspects:

\begin{figure}[t]
  \centering
  \begin{minipage}[t]{0.48\linewidth}
    \centering
    \includegraphics[width=\linewidth]{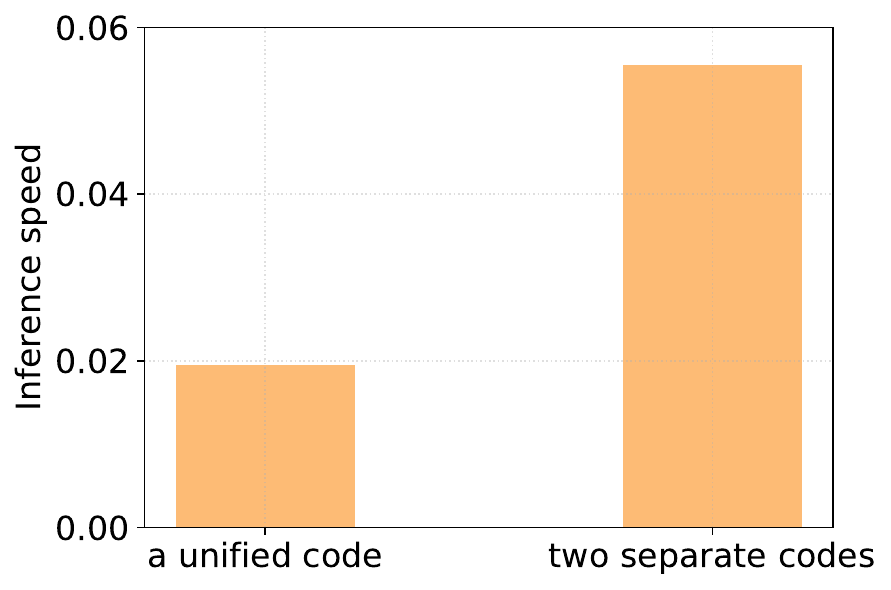}
    \caption{Comparison of inference speed (second per sample, topk=5, beam size=100) between a unified code and two separate codes setups on the Beauty dataset.}
    \label{fig:speed}
  \end{minipage}%
  \hfill
  \begin{minipage}[t]{0.48\linewidth}
    \centering
    \includegraphics[width=\linewidth]{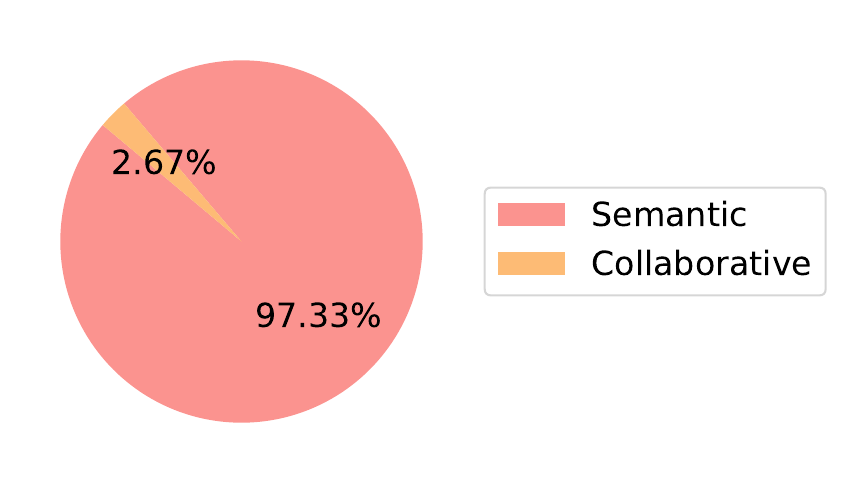}
    \caption{Proportional similarity of semantic modality and collaborative modality to the final representation with concatenation method on the Beauty dataset.}
    \label{fig:concat}
  \end{minipage}
\end{figure}

Firstly, \textbf{a unified code} that encodes both collaborative and semantic knowledge is crucial for practical deployment. In real-world recommendation systems, storing and retrieving items efficiently requires a compact and discrete representation that integrates all essential information related to items. However, due to the inherent differences between collaborative and semantic modalities \cite{kim2024sc, wang2024enhanced}, directly combining them often leads to suboptimal representations. Semantic embeddings derived from large language models typically contain rich contextual signals, while collaborative embeddings reflect users' actual preferences through interaction histories. To this end, we propose a two-stage knowledge integration framework. In the first stage, we separately extract modality-specific knowledge using two pretrained models. Specifically, we adopt Llama2-7b \cite{touvron2023llama} to obtain semantic representations based on textual descriptions and DIN \cite{zhou2018deep} to model collaborative behavior patterns. These modality-specific embeddings are then fused through a modality-adaptive fusion module, which is designed to learn an integrated representation that balances and aligns the semantic and collaborative knowledge. The integrated embeddings are subsequently quantized into discrete item codes, referred to as \textit{Unicodes}, which act as universal identifiers for each item in our model. In the second stage, we utilize the Unicodes within a sequence modeling architecture for recommendation. An encoder is employed to capture the user’s interaction history, while a decoder predicts the Unicode corresponding to the next item. This reformulation allows the recommendation task to be viewed as a discrete sequence generation task.

Secondly, to \textbf{resolve the semantic domination issue}, we introduce two components: a learnable modality adaptation layer and a cross-modality knowledge alignment task. The adaptation layer adjusts the relative emphasis of each modality during the fusion process, allowing the model to modulate their contributions dynamically. Meanwhile, the knowledge alignment task provides explicit supervision by aligning the distributions of semantic and collaborative embeddings through a contrastive learning objective. By encouraging the representations of the same item across modalities to be close in the latent space, the model is guided to capture shared semantics while preserving modality-specific features. This alignment is jointly optimized with the next-item prediction objective, allowing the model to benefit from both modality alignment and end-task supervision. This joint training encourages a more balanced representation that leverages complementary knowledge and reduces the risk of over-reliance on semantic features.


Thirdly, ensuring \textbf{comprehensive and sufficient learning} is essential for fully exploiting both knowledge. The quantization process, although necessary for obtaining discrete Unicodes, inevitably introduces approximation errors that may lead to partial information loss. Since the quantized codes are derived by approximating the original embeddings, relying solely on them may result in insufficient representation of the items. To compensate for this, we introduce an auxiliary learning objective based on intra-modality knowledge distillation, where the original embeddings are leveraged to provide more complete and informative guidance.
Motivated by the effectiveness of the [CLS] token in BERT \cite{devlin2018bert}, we incorporate a special token into the input sequence during the second stage. This token serves to summarize the sequence's informative content. We then conduct a global contrastive learning task between the embedding of this token and the integrated embedding of the target item. This distillation task enables the model to capture and retain high-level modality-specific knowledge that might be attenuated during quantization. By maximizing agreement between the sequence summary and the target item’s unified representation, the model learns to distill and refine knowledge more effectively.

In summary, our contributions can be concluded as follows:
\begin{itemize}
\item We introduce UNGER, with a unified code, Unicodes, which integrates both collaborative and semantic knowledge for generative recommendation. By utilizing the unified code, we achieve faster inference while requiring only half the storage space compared to two separate codes setup.
\item We propose a learnable adaption layer and a joint optimization of cross-modality knowledge alignment and next item prediction tasks to adaptively learn an integrated embedding for resolving the semantic domination issue, and introduce an intra-modality knowledge distillation task to ensure comprehensive and sufficient learning, compensating for the information loss.
\item Extensive experiments on three public recommendation benchmarks demonstrate UNGER’s superiority over existing methods, encompassing both generative and traditional paradigms. Additionally, we observe that UNGER demonstrates the scaling law characteristic.
\end{itemize}

\section{RELATED WORK}
In this section, we review the relevant literature and highlight the distinctions between our approach and the existing studies.

\subsection{Pre-training in Recommender Systems}
 Similar to the field of Natural Language Processing (NLP), the advancement of recommendation models also encounters challenges due to the data sparsity \cite{pan2019warm}. Traditional recommendation approaches depend heavily on extensive user–item interaction records to model both user preferences and item attributes, as well as the latent relationships between users and items, but when new users or new items come in with few or no interaction histories, these methods are struggle to model their latent representations accurately, resulting in poor recommendation quality and giving rise to the \textit{cold-start problem} in recommender systems \cite{lika2014facing, lam2008addressing, li2025personalized}, which significantly restricts the scalability and efficacy of classical recommendation paradigms. 

To address this challenge, researchers have increasingly turned to pre-training strategies inspired by breakthroughs in NLP. Early methods primarily focus on learning universal user representations from large-scale interaction data, often using self-supervised or multi-task learning objectives based on ID-level signals. For example, PeterRec \cite{yuan2020parameter} enables knowledge transfer across multiple downstream tasks through parameter-efficient fine-tuning, while Conure \cite{yuan2021one} introduces a lifelong learning framework that incrementally updates user embeddings without forgetting previous knowledge. CLUE \cite{cheng2021learning} further improves representation quality by adopting contrastive learning at the sequence level, bridging the gap between item-level objectives and user-level semantics. However, such ID-based approaches struggle with transferability when applied to new domains or scenarios, as they rely heavily on discrete user and item identifiers that lack generalization capability. To address this, several work incorporate rich side information such as item descriptions and user reviews to enhance semantic understanding. ZESRec \cite{ding2021zero} replaces item IDs with textual metadata to enable zero-shot generalization across entirely disjoint domains, while UniSRec \cite{hou2022towards} and MISSRec \cite{wang2023missrec} leverage multimodal features including text and images to build transferable sequence and item representations that better reflect user preferences. Building upon these advances, P5 \cite{geng2022recommendation} unifies various recommendation tasks into a text-to-text framework, transforming all inputs and outputs into natural language and enabling a single language model to handle diverse recommendation objectives with minimal adaptation. Nevertheless, these approaches still rely on small-scale pretrained language models, which constrains their capacity to capture universal world knowledge and provide general representation.

\subsection{Sequential Recommendation}
 Sequential recommendation models focus on modeling user behavior as a chronologically ordered sequence of interactions, aiming to predict the next item a user will engage with.\\
{\bfseries Traditional Approaches.} Early approaches primarily utilized Markov Chains (MCs)  \cite{rendle2010factorizing, he2016fusing} to capture the transition probabilities between items. With the rise of deep learning, an ID-based recommendation paradigm emerged. Various deep neural network architectures have been developed under this paradigm. For instance, GRU4Rec \cite{jannach2017recurrent} was the first to employ GRU-based RNNs for sequential recommendations, while SASRec \cite{kang2018self} introduced a self-attention mechanism similar to decoder-only transformer models, to capture long-range dependencies. Inspired by the success of masked language modeling, BERT4Rec \cite{sun2019bert4rec} utilized transformers with masking strategies to enhance sequential recommendation tasks. Building on the masking technique, S\textsuperscript{3}-Rec \cite{zhou2020s3} learns the correlations among attributes, items, subsequences, and sequences through the principle of mutual information maximization (MIM), thereby enhancing data representation.

\noindent{\bfseries Generative Approaches.} Unlike traditional embedding-based methods, which rely on dot-product (cosine) similarity and external ANN search systems for top-k retrieval, generative approaches predict item identifiers directly. 
Generative methods can be broadly categorized into two types: prompt fine-tuning strategies based on \textit{off-the-shelf} large language models (LLMs) and training from scratch for custom-designed models.

For LLMs based methods \cite{zheng2024adapting, lin2024bridging, liu2024collaborative, shen2024pmg, kim2024sc, ji2024genrec}, the focus is on designing refined prompts and fine-tuning tasks that help language models better understand recommendation tasks. LC-Rec \cite{zheng2024adapting} introduces a learning-based vector quantization approach for semantically meaningful item indexing and fine-tuning tasks that align collaborative signals with LLM representations, achieving superior performance in diverse scenarios. CCF-LLM \cite{liu2024collaborative} transforms user-item interactions into hybrid prompts encoding both semantic knowledge and collaborative signals, utilizing an attentive cross-modal fusion strategy to integrate embeddings from different modalities. SC-Rec \cite{kim2024sc} utilizes multiple item indices and prompt templates, alongside a self-consistent re-ranking mechanism, to more effectively merge collaborative and semantic knowledge.

For methods that train from scratch \cite{zeng2024scalable, wang2024enhanced, liu2024mmgrec, feng2022recommender, rajput2023recommender, wang2024eager, zhu2018learning, zhu2019joint, wang2024learnable}, the primary focus is on converting the raw sequence recommendation task into an autoregressive generation task. Tree-based methods \cite{zhu2018learning, feng2022recommender, zhu2019joint}, such as RecForest  \cite{feng2022recommender}, have shown promising performance by constructing multiple trees and integrating transformer-based structures for routing. Additionally, TIGER \cite{rajput2023recommender} introduced the concept of semantic IDs, representing each item as a set of tokens derived from its side information, and predicting next item tokens in a sequence-to-sequence manner. EAGER \cite{wang2024eager} employs a dual-stream generative framework to parallely utilize semantic and behavioral information with two separate codes, generating recommended items from each respective pipeline, and ultimately selecting the top-k items based on confidence scores. 

In this paper, we aim to seamlessly integrate collaborative and semantic modality knowledge into a unified code for generative recommendation, while ensuring no additional computational and storage overhead compared with solely utilizing one modality.

\section{METHODOLOGY}

\subsection{Problem Formulation}
Given an item corpus \( I \) and a user’s historical interaction sequence \( U = [u_1, u_2, \dots, u_{t-1}] \) where \( u \in I \), the target of a sequential recommendation system is to predict  the next most likely item \( u_t \in I \) that the user may interact with.

In the generative framework, each item \( u \) is represented by a sequence of codes \( C = [c_1, c_2, \dots, c_L] \), where \( L \) denotes the length of the code sequence. Here, the sequential recommendation task shifts to predicting the codes \( C_t \) of the next item \( u_t \) based on the user's historical interaction sequence \( U \). During training, the model first encodes the user’s historical interaction sequence \( U \) and then generates the codes \( C_t \) of the target item \( u_t \) step by step at the decoder. The decoding process is defined by the following formula:
\begin{equation}
p(C_t|U) = \prod_{i=1}^{L} p(c_i|U, c_1, c_2, \dots, c_{i-1})
\end{equation}
During the inference phase, the decoder performs beam search to autoregressively generate the codes of the top-k items.

\begin{figure*}[htbp]
  \centering
  \includegraphics[width=\linewidth]{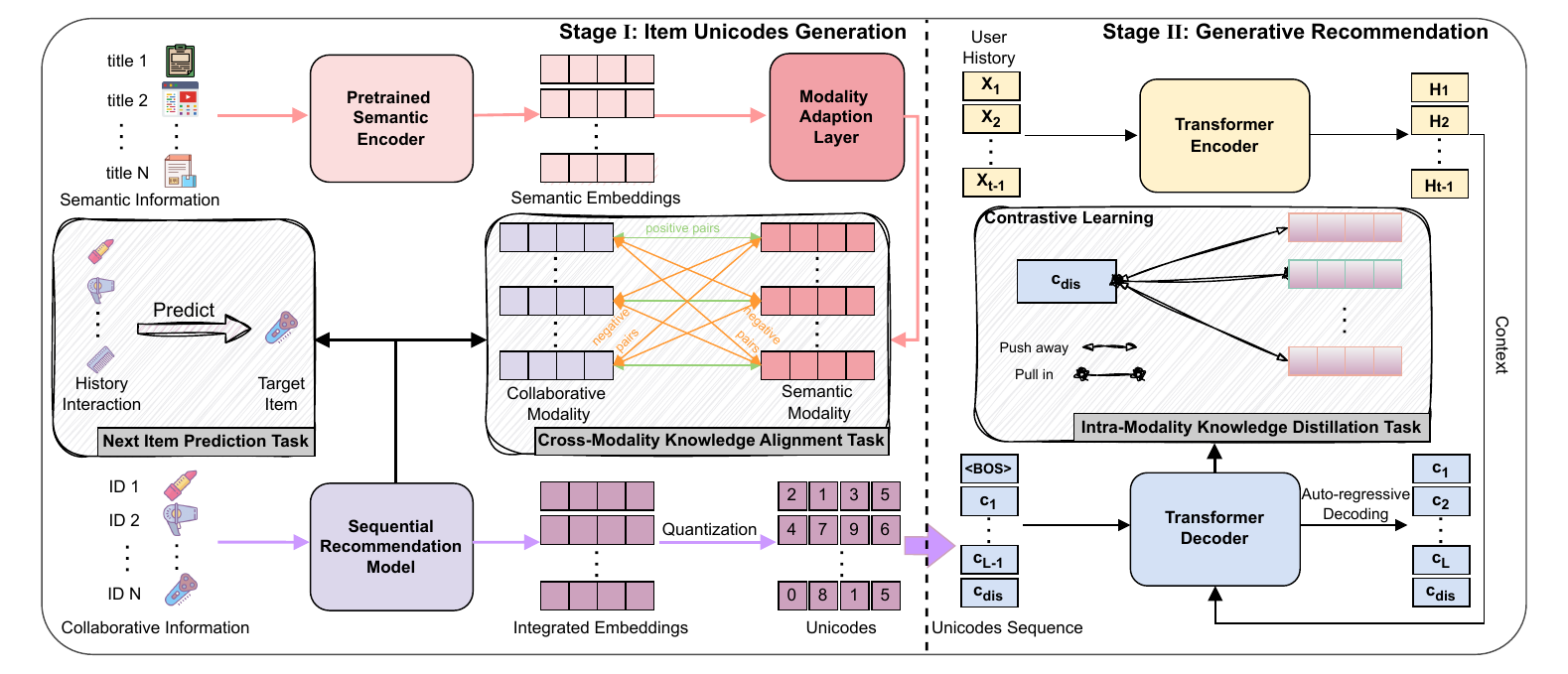}
  \caption{An overview of UNGER. UNGER consists of two stages. The first stage integrates semantic and collaborative knowledge to construct the unified code, Unicode for each item. The second stage utilizes the obtained unicodes to perform generative recommendation. To achieve the goal of utilizing a unified code to encode the two different knowledge for generative recommendation, we introduce two auxiliary tasks at each stage: a cross-modality knowledge alignment task (CKA) in the first stage and an intra-modality knowledge distillation task (IKD) in the second stage. Besides, in the first stage, a modality adaption layer with AdaLN is also introduced to bridge the modality gap between semantic and collaborative space.}
  \label{fig:UNGER}
\end{figure*}

\subsection{Overall Pipeline}
We present UNGER overall framework in Figure \ref{fig:UNGER}, which consists of two stages. In the first stage, we first adaptively learn integrated embeddings through the joint optimization of proposed cross-modality knowledge alignment task and next item prediction task, which fuse knowledge from both collaborative and semantic modalities. Subsequently, the integrated embeddings are quantized into discrete unicodes that act as item identifiers in the second stage. The second stage is the generative recommendation phase, comprising an encoder, a decoder and an intra-modality knowledge distillation task to compensate for the information loss caused by the quantization process. Here, the user’s historical interaction sequence is transformed into unicodes sequence using the item-unicode mapping table obtained in the first stage. The encoder models the user's historical interactions to capture their interests, which are then passed into the decoder to guide the generation of unicodes for candidate items. Once training is complete, we select the top-\(k\) recommendations based on confidence scores.

\subsection{Stage \uppercase\expandafter{\romannumeral 1}: Item Unicodes Generation}
In this stage, our objective is to construct item unicodes which are capable of encoding both collaborative and semantic knowledge while utilizing a single unified codebook. Let \( X \) and \( S \) denote the ID sequence and corresponding item semantic side information (e.g. such as titles) of the user’s historical interaction sequence $U$, respectively. Then, we first use a randomly initialized sequential recommendation model (e.g. DIN \cite{zhou2018deep}) to encode the user’s historical sequence \( X \) into collaborative embeddings \( E_C \), and employ a pretrained semantic encoder (e.g. Llama2-7b \cite{touvron2023llama}) to encode items’ semantic knowledge \( S \) into embeddings \( E_S \). 

\subsubsection{Modality adaption layer} To bridge the modality gap between two modalities, we propose using a learnable modality adaptation layer with AdaLN that maps the semantic embeddings \( E_S \) into the same embedding space as collaborative embeddings. Unlike standard Layer Normalization, AdaLN \cite{peebles2023scalable} introduces learnable affine parameters that are conditioned on the input itself, enabling the model to dynamically adjust the normalization process based on the characteristics of each modality. This design choice is particularly well-suited for our setting, where semantic and collaborative embeddings inherently capture distinct patterns (e.g., content-driven semantics vs. behavior-driven preferences). By leveraging AdaLN, the adaptation layer is better equipped to preserve and align these modality-specific signals. The mapping process is defined as follows: 


\begin{equation}
E_T = \text{MAL}(E_S) = \text{AdaLN}(W E_S + b) 
\end{equation}

where MAL refers to the modality adaptation layer, with \( W \) and \( b \) as learnable parameters, and \text{AdaLN} indicating the adaptive normalization layer.

\subsubsection{Cross-modality knowledge alignment task} To effectively integrate semantic and collaborative knowledge, we introduce a cross-modality knowledge alignment task designed to align embeddings from the collaborative modality \( E_C \) and semantic modality \( E_T \). For a given item \( i \in I\), our objective is to bring its collaborative embedding \( E_{C_i} \) closer to its semantic embedding \( E_{T_i} \) for item \( i \) while distancing \( E_{C_i} \) from the semantic embedding \( E_{T_j} \) of a different item \( j \in I\), where \( j \) is an in-batch negative sample selected from the same batch as item \( i \). By structuring the alignment in this way, the learned embedding for item \( i \) is able to encapsulate knowledge from both modalities. To achieve this goal, we adopt the Info-NCE loss \cite{chen2020simple}. The loss function for our cross-modality knowledge alignment task is defined as follows:

\begin{equation}\label{eq:alignLoss}
\mathcal{L}_{\text{align}} = -\log {\sum_{i \in I}} \frac{\exp(\text{sim}(E_{C_i}, E_{T_i}) / \tau)}{\sum\limits_{{\substack{j \in \text{in-batch neg-samples} \\ j \neq i}}} \exp(\text{sim}(E_{C_i}, E_{T_j}) / \tau)}
\end{equation}

where \(\text{sim}(\cdot, \cdot)\) represents a similarity function between embeddings (e.g. dot product or cosine similarity), and \( \tau \) is a temperature parameter used to control the smoothness of the similarity.

\subsubsection{Next item prediction task} In addition to the alignment loss described above, we adopt next item prediction task to enhance the model’s ability to  capture the sequence dependency, thereby supporting the learning of robust collaborative representations. This task takes the user’s historical interaction sequence \([ x_1, x_2, \dots, x_{t-1} ]\) as input, learns a representation of the user's preferences, and computes matching scores by measuring the similarity between the learned preferences and candidate items. These scores are subsequently normalized to derive the probability distribution over the target item. The goal is to maximize the probability assigned to the actual target item \( x_t \), which can be formally represented as follows:
\begin{equation}\label{eq:seqLoss}
\mathcal{L}_{\text{seq}} = - \sum_{t=2}^{L} \log p(x_t \mid x_1, x_2, \dots, x_{t-1})
\end{equation}

\subsubsection{Joint optimization} In the first stage, we jointly optimize the two losses to train the modality adaption layer and sequential recommendation model. The total loss function for the first stage is defined as follows:
\begin{equation}
\label{stage1}
\mathcal{L}_{\text{Stage \uppercase\expandafter{\romannumeral 1}}} = \mathcal{L}_{\text{seq}} + \alpha \mathcal{L}_{\text{align}}
\end{equation}

where $\mathcal{L}_{\text{seq}}$ is the next item prediction loss defined in equation (\ref{eq:alignLoss}), $\mathcal{L}_{\text{align}}$ is the cross-modality knowledge alignment loss defined in equation (\ref{eq:seqLoss}), and \( \alpha \) is a tunable hyperparameter used to adjust the relative importance of the alignment loss.


\subsubsection{Unified Codes} 

After the training is finished, following previous work~\cite{wang2024eager}, we apply the hierarchical k-means algorithm through an iterative residual quantization process. As depicted in Algorithm \ref{alg:hierarchical_kmeans}, this method allows us to efficiently encode high-dimensional item embeddings into discrete hierarchical codes, while preserving as much information as possible at each stage of quantization. The core mathematical relationship at each layer is defined as:

\begin{equation}
    \mathbf{r}_i^l = \mathbf{r}_i^{l-1} - \mathbf{C}_i^l
\end{equation}

where $\mathbf{r}_i^0 = \mathbf{v}_i$ denotes the original embedding of item $i$, and $\mathbf{C}_i^l$ represents the centroid assigned to item $i$ at the $l$-th layer of quantization. This recursive structure ensures that each subsequent residual $\mathbf{r}_i^l$ captures the portion of the embedding not explained by the current centroid approximation.

\begin{algorithm}[ht]
    \caption{Hierarchical K-means Clustering}
    \label{alg:hierarchical_kmeans}
    \begin{algorithmic}[0]
        \Statex \textbf{Input:} Item embeddings $\mathcal{V} = \{\mathbf{v}_1, \mathbf{v}_2, \dots\}$, number of clusters $K$, hierarchy depth $L$
        \Statex \textbf{Output:} Unicode sequences $\mathbf{c}_i = [c_i^1, c_i^2, \dots, c_i^L]$ for all items
    \end{algorithmic}
    \begin{algorithmic}[1]
        \For{each item $i$}
            \State Initialize residual: $\mathbf{r}_i^0 \gets \mathbf{v}_i$
        \EndFor
        \For{layer $l \gets 1$ to $L$}
            \State Collect residuals: $\mathcal{R}^{l-1} \gets \{\mathbf{r}_1^{l-1}, \mathbf{r}_2^{l-1}, \dots\}$
            \State Perform k-means clustering on $\mathcal{R}^{l-1}$ with $K$ clusters
            \State Store centroids in codebook $C_l = \{\mathbf{C}_1^l, \mathbf{C}_2^l, \dots, \mathbf{C}_K^l\}$
            \For{each item $i$}
                \State Compute distances: $d_k = \|\mathbf{r}_i^{l-1} - \mathbf{C}_k^l\|^2$ for all $k \in [1,K]$
                \State Assign cluster index: $c_i^l \gets \arg\min_{k} d_k$
                \State Record centroid: $\mathbf{C}_i^l \gets \mathbf{C}_{c_i^l}^l$
                \State Update residual: $\mathbf{r}_i^l \gets \mathbf{r}_i^{l-1} - \mathbf{C}_i^l$
            \EndFor
        \EndFor
    \end{algorithmic}
\end{algorithm}

The algorithm proceeds as follows: We begin by initializing the residual vectors with the original item embeddings, i.e., $\mathbf{r}_i^0 = \mathbf{v}_i$. At each hierarchical layer $l$, we collect all current residuals $\mathcal{R}^{l-1}$ and apply k-means clustering to partition them into $K$ clusters. This step identifies common patterns or structures in the embedding space that can be captured by a shared centroid.

The set of $K$ resulting centroids is stored as the codebook $C_l = \{\mathbf{C}_1^l, \mathbf{C}_2^l, \dots, \mathbf{C}_K^l\}$. For every item $i$, we compute the squared Euclidean distance between its current residual $\mathbf{r}_i^{l-1}$ and each centroid in $C_l$:

\begin{equation}
    d_k = \|\mathbf{r}_i^{l-1} - \mathbf{C}_k^l\|^2
\end{equation}

The nearest centroid is then selected based on the minimum distance criterion:

\begin{equation}
    c_i^l = \arg\min_{k} d_k
\end{equation}

Once the closest centroid $\mathbf{C}_i^l = \mathbf{C}_{c_i^l}^l$ is determined, it is subtracted from the current residual to compute the new residual for the next layer:

\begin{equation}
    \mathbf{r}_i^l = \mathbf{r}_i^{l-1} - \mathbf{C}_i^l
\end{equation}

This subtraction removes the information captured by the current centroid, allowing the next layer to focus on the remaining details. This process is repeated iteratively across $L$ layers, with each layer progressively refining the approximation of the original embedding by capturing finer-grained residuals.

After all $L$ layers have been processed, we obtain a sequence of discrete cluster indices $\mathbf{c}_i = [c_i^1, c_i^2, \dots, c_i^L]$ for each item. This sequence serves as a compact and hierarchical unicode representation of the item embedding. The hierarchical structure allows for efficient indexing and retrieval while preserving high fidelity in the quantization process, as each level incrementally refines the representation by focusing on residuals not explained by previous levels. 

Subsequently, we construct an item-unicode lookup table that maps each item to its corresponding unicode sequence. These unicodes serve as discrete identifiers for the items and play a crucial role in Stage \uppercase\expandafter{\romannumeral 2} of our framework, where they are used to facilitate generative recommendation. By operating in the unicode space, the model can  manipulate item representations in a structured and semantically meaningful manner, enabling flexible and efficient recommendation generation.


\subsection{Stage \uppercase\expandafter{\romannumeral 2}: Generative Recommendation}

\subsubsection{Encoding process} On the encoder side, given a user’s interaction history \( X = [x_1, x_2, \dots, x_{t-1}] \), we input it into an encoder consisting of stacked multi-head self-attention layers and feed-forward layers, following the Transformer architecture. The encoder processes \( X \) to produce a feature representation \( H \), which captures the user's interests and will be passed to the decoder.

\subsubsection{Decoding process} On the decoder side, items are mapped to their unicodes derived in Stage  \uppercase\expandafter{\romannumeral 1}. Our objective changes from predicting the next item \( x_t \) to predicting its item unicode \( [c_1, c_2, \dots, c_L] \). For training, a special token \texttt{<BOS>} is prepended to the item unicode to construct the decoder input. The generative recommendation loss is computed using a cross-entropy loss function, defined as follows:
\begin{equation}
\mathcal{L}_{\text{gen}} = \sum_{i=1}^L \log p(c_i \mid x, \texttt{<BOS>}, c_1, \dots, c_{i-1}),
\end{equation}

\subsubsection{Intra-modality knowledge distillation task} To compensate for the information loss caused by the quantization process in the first stage, we introduce an intra-modality knowledge distillation task to distill the global knowledge. Inspired by the use of the [CLS] token in BERT for capturing global context in classification tasks, we append a learnable token $[c_{\text{dis}}]$ to the end of the decoder input, allowing gradients to propagate through each preceding unicode. This design encourages $[c_{\text{dis}}]$ to capture global information.

Specifically, the final layer output corresponding to $[c_{\text{dis}}]$ is used in a global contrastive learning objective over the item corpus: the positive sample is the integrated embedding $E_t$ of item $x_t$ learned from the first stage, and the negative sample is a randomly selected integrated embedding $E_{\text{neg}}$ from other items in the corpus, excluding $x_t$.
This objective pulls the $[c_{\text{dis}}]$ final layer output closer to the positive sample $E_t$ and pushes it away from the negative sample $E_{\text{neg}}$. The loss function for the distillation task is defined as follows:
\begin{equation}
\mathcal{L}_{\text{distillation}} = -\log \frac{\exp(c_{\text{dis}} \cdot E_t )}{\exp(c_{\text{dis}} \cdot E_t ) + \sum \exp(c_{\text{dis}} \cdot E_{\text{neg}} )}
\end{equation}

where \( c_{\text{dis}} \) denotes the final layer output of the decoder for the special token \([c_{\text{dis}}]\); \( E_{\text{neg}} \) is the integrated embedding for a negative sample as learned in Stage \uppercase\expandafter{\romannumeral 1}, while \( E_{t} \) represents the integrated embedding of the target item, also learned in Stage \uppercase\expandafter{\romannumeral 1}.

\subsubsection{Joint optimization} Finally, in Stage \uppercase\expandafter{\romannumeral 2}, we jointly optimize these two loss functions. The overall loss function for Stage \uppercase\expandafter{\romannumeral 2} is defined as follows:

\begin{equation}
\label{stage2}
\mathcal{L}_{\text{Stage \uppercase\expandafter{\romannumeral 2}}} = \mathcal{L}_{\text{gen}} + \beta \mathcal{L}_{\text{distillation}}
\end{equation}

where $\mathcal{L}_{\text{gen}}$ is the generative recommendation loss, $\mathcal{L}_{\text{distillation}}$ is the intra-modality knowledge distillation loss, and $\beta$ is a hyperparameter that balances the two objectives.

\subsection{Training and Inference}
\subsubsection{Training} we employ a two-stage training process. In the first stage, we use DIN as the backbone and optimize the model with loss \( \mathcal{L}_{\text{stage \uppercase\expandafter{\romannumeral 1}}} \). After the first stage training is complete, we extract the item integrated embeddings and derive item unicodes. In the second stage, we use a Transformer for generative recommendation, optimizing it with loss \( \mathcal{L}_{\text{stage \uppercase\expandafter{\romannumeral 2}}} \).

\subsubsection{Inference} During inference phase, we rely exclusively on the Transformer model. In the decoding process, beam search is applied iteratively to generate each token within the item unicode sequence. Once finished, the item unicodes are mapped to their corresponding items through the item-unicode lookup table, creating a ranked recommendation list based on confidence scores to produce the final top-$k$ results.

\subsection{Computational and Storage Costs Analysis}


In this section, we analyze the computational and storage efficiency of our proposed method compared to existing approaches that handle multiple modalities separately. Our method leverages a unified code to jointly encode both semantic and collaborative information, which offers significant advantages in terms of both computational complexity and memory usage.

\subsubsection{Computational Cost}
In our framework, the encoding phase processes the user interaction sequence only once through the encoder, which yields a latent representation. This one-time encoding step ensures that the computational overhead during encoding remains minimal. The main computational burden lies in the decoding stage, where the decoder autoregressively generates each token of the target item code. Assuming the code length is fixed and the decoding of each item code takes $\mathcal{O}(n)$ time, the total decoding complexity for generating a recommendation is $\mathcal{O}(n)$.

In contrast, existing methods such as \cite{wang2024eager, kim2024sc} adopt separate modality-specific codes, typically maintaining $K$ parallel decoders—each corresponding to a different modality (e.g., semantic, collaborative, etc.). To generate recommendations, these methods must decode $K$ modality-specific codes in parallel, leading to a cumulative decoding cost of $\mathcal{O}(Kn)$. Moreover, these approaches usually include an additional ranking or fusion step based on confidence scores to integrate the outputs from different modalities. While we exclude the computational cost of this post-decoding ranking step in our comparison, it is worth noting that it may further increase the overall latency during inference.

In summary, by unifying all modalities into a single code and decoding only once, our approach significantly reduces the computational complexity from $\mathcal{O}(Kn)$ to $\mathcal{O}(n)$, offering a more scalable solution especially when the number of modalities $K$ is large.

\subsubsection{Storage Cost}
Let us now consider the memory requirements for storing the item codes. Suppose that storing one modality-specific code requires $\mathcal{O}(m)$ space. For methods that encode each modality independently, storing $K$ codes per item entails a total storage cost of $\mathcal{O}(Km)$. 
By contrast, our method compresses all relevant information into a single unified code, thus requiring only $\mathcal{O}(m)$ storage per item. This compression not only reduces the memory footprint but also simplifies the storage architecture and retrieval process during inference.

Overall, our unified coding strategy delivers substantial improvements in both computational and storage efficiency. The reduction from $\mathcal{O}(Kn)$ to $\mathcal{O}(n)$ in decoding time, and from $\mathcal{O}(Km)$ to $\mathcal{O}(m)$ in storage requirements, makes our approach well-suited for large-scale recommender systems where real-time performance and memory scalability are critical.

\begin{table}[htbp]
    \centering
    \caption{Comparison of our method with typical generative recommendation methods TIGER and EAGER on computational and storage costs.}
    \label{tab:costs analysis}
    \resizebox{\linewidth}{!}{
    \begin{tabular}{cccc}
        \toprule
         &  TIGER & EAGER & UNGER (Ours)\\
        \midrule
        Computation Cost & $\mathcal{O}(n)$ & $\mathcal{O}(2n)$ & $\mathcal{O}(n)$ \\
        Storage Cost     & $\mathcal{O}(m)$ & $\mathcal{O}(2m)$ & $\mathcal{O}(m)$ \\
        Used Modality   & Semantic Only & Semantic + Collaborative &  Semantic + Collaborative \\
        \bottomrule
    \end{tabular}
    }
\end{table}

\begin{table}[htbp]
    \centering
    \caption{Statistics of the Datasets.}
    \label{tab:Statistics of the Datasets}
    \resizebox{0.7\linewidth}{!}{
    \begin{tabular}{ccccc}
        \toprule
        Dataset & \#Users & \#Items & \#Interactions & \#Density \\
        \midrule
        Beauty & 22,363 & 12,101 & 198,360 & 0.00073 \\
        Sports and Outdoors & 35,598 & 18,357 & 296,175 & 0.00045 \\
        Toys and Games & 19,412 & 11,924 & 167,526 & 0.00073 \\
        \bottomrule
    \end{tabular}
    }
\end{table}

\section{EXPERIMENTS}
We analyze the proposed UNGER method and demonstrate its effectiveness by answering the following research questions:

\begin{itemize}[left=0pt]
\item \textbf{RQ1}: How does UNGER perform compared with existing best-performing sequential recommendation methods among different datasets?
\item \textbf{RQ2}: How effective is UNGER in mitigating the semantic domination issue?
\item \textbf{RQ3}: Do cross-modality knowledge alignment and intra-modality knowledge distillation tasks each contribute positively to UNGER’s performance?
\item \textbf{RQ4}: How do different ablation variants and hyper-parameter settings affect the performance of UNGER?
\end{itemize}

\subsection{Experimental Setting}
\subsubsection{Dataset} We conduct experiments on three public benchmarks commonly used in the sequential recommendation task. For all datasets, we group the interaction records by users and sort them by the interaction timestamps ascendingly. Following \cite{rendle2010factorizing,zhang2019feature}, we only keep the 5-core dataset, which filters unpopular items and inactive users with fewer than five interaction records. Statistics of these datasets are shown in Table \ref{tab:Statistics of the Datasets}.
\begin{itemize}
\item \textbf{Amazon}: Amazon Product Reviews dataset \cite{mcauley2015image}, containing user reviews and item metadata from May 1996 to July 2014. In particular, we use three categories of the Amazon Product Reviews dataset for the sequential recommendation task: "Beauty", "Sports and Outdoors", and "Toys and Games".
\end{itemize}

\begin{table*}[htbp]
    \renewcommand{\arraystretch}{1.5}
    \centering
    \caption{Performance comparison of different methods. The best performance is highlighted in bold while the second best performance is underlined. All the results of UNGER are statistically significant with paired t-test $p < 0.05$ compared to the best baseline models. We computed the standard error for all metrics for the datasets by running the training and evaluation 5 times using different random seeds chosen in [2020, 2021, 2022, 2023, 2024].}
    \label{tab:overall performance}
    \resizebox{\linewidth}{!}{
    \begin{tabular}{llcccccccccccc}
        \toprule
        \multirow{2}{*}{\textbf{Dataset}} & \multirow{2}{*}{\textbf{Metric}} & \multicolumn{7}{c}{\textbf{Classical}} & \multicolumn{4}{c}{\textbf{Generative}} & \multirow{2}{*}{\textbf{UNGER (Ours)}} \\
        \cmidrule(lr){3-9} \cmidrule{10-13}
        & & \textbf{GRU4REC} & \textbf{Caser} & \textbf{SASRec} & \textbf{BERT4Rec} & \textbf{HGN} & \textbf{FDSA} & \textbf{S$^3$-Rec} & \textbf{Recorest} & \textbf{TIGER} & \textbf{ColaRec} & \textbf{EAGER} \\
        \midrule
        \multirow{4}{*}{\textbf{Beauty}} 
        & Recall@10 & 0.0283 & 0.0347 & 0.0605 & 0.0347 & 0.0512 & 0.0407 & 0.0647 & 0.0664 & 0.0617\textsuperscript{*} & 0.0524\textsuperscript{*} & \underline{0.0836} & \textbf{0.0939} $\pm$ 0.00093 \\
        & Recall@20 & 0.0479 & 0.0556 & 0.0902 & 0.0599 & 0.0773 & 0.0656 & 0.0994 & 0.0915 & 0.0924\textsuperscript{*} & 0.0807\textsuperscript{*} & \underline{0.1124} & \textbf{0.1289} $\pm$ 0.00084 \\
        & NDCG@10 & 0.0137 & 0.0176 & 0.0318 & 0.0170 & 0.0266 & 0.0208 & 0.0327 & 0.0400 & 0.0339\textsuperscript{*} & 0.0263\textsuperscript{*} & \underline{0.0525} & \textbf{0.0559} $\pm$ 0.00037 \\
        & NDCG@20 & 0.0187 & 0.0229 & 0.0394 & 0.0233 & 0.0332 & 0.0270 & 0.0414 & 0.0464 & 0.0417\textsuperscript{*} & 0.0335\textsuperscript{*} & \underline{0.0599} & \textbf{0.0646} $\pm$ 0.00041 \\
        \midrule
        \multirow{4}{*}{\textbf{Sports}} 
        & Recall@10 & 0.0204 & 0.0194 & 0.0350 & 0.0191 & 0.0313 & 0.0288 & 0.0385 & 0.0247 & 0.0376\textsuperscript{*} & 0.0348\textsuperscript{*} & \underline{0.0441} & \textbf{0.0471} $\pm$ 0.00078 \\
        & Recall@20 & 0.0333 & 0.0314 & 0.0507 & 0.0315 & 0.0477 & 0.0463 & 0.0607 & 0.0375 & 0.0577\textsuperscript{*} & 0.0533\textsuperscript{*} & \underline{0.0659} & \textbf{0.0710} $\pm$ 0.00075 \\
        & NDCG@10 & 0.0110 & 0.0097 & 0.0192 & 0.0099 & 0.0159 & 0.0156 & 0.0204 & 0.0133 & 0.0196\textsuperscript{*} & 0.0179\textsuperscript{*} & \underline{0.0236} & \textbf{0.0259} $\pm$ 0.00034 \\
        & NDCG@20 & 0.0142 & 0.0126 & 0.0231 & 0.0130 & 0.0201 & 0.0200 & 0.0260 & 0.0164 & 0.0246\textsuperscript{*} & 0.0226\textsuperscript{*} & \underline{0.0291} & \textbf{0.0319} $\pm$ 0.00049 \\
        \midrule
        \multirow{4}{*}{\textbf{Toys}} 
        & Recall@10 & 0.0176 & 0.0270 & 0.0675 & 0.0203 & 0.0497 & 0.0381 & 0.0700 & 0.0383 & 0.0578\textsuperscript{*} & 0.0474\textsuperscript{*} & \underline{0.0714} & \textbf{0.0822} $\pm$ 0.00085 \\
        & Recall@20 & 0.0301 & 0.0420 & 0.0941 & 0.0358 & 0.0716 & 0.0632 & 0.1065 & 0.0483 & 0.0838\textsuperscript{*} & 0.0704\textsuperscript{*} & \underline{0.1024} & \textbf{0.1154} $\pm$ 0.00070 \\
        & NDCG@10 & 0.0084 & 0.0141 & 0.0374 & 0.0099 & 0.0277 & 0.0189 & 0.0376 & 0.0285 & 0.0321\textsuperscript{*} & 0.0242\textsuperscript{*} & \underline{0.0489} & \textbf{0.0505} $\pm$ 0.00032 \\
        & NDCG@20 & 0.0116 & 0.0179 & 0.0441 & 0.0138 & 0.0332 & 0.0252 & 0.0468 & 0.0310 & 0.0386\textsuperscript{*} & 0.0300\textsuperscript{*} & \underline{0.0538} & \textbf{0.0573} $\pm$ 0.00025 \\
        \bottomrule
    \end{tabular}}
\end{table*}

\subsubsection{Evaluation Metrics} We employ two broadly used criteria for the matching phase, \textit{i.e.}, Recall and Normalized Discounted Cumulative Gain (NDCG). We report metrics computed on the top 10/20 recommended candidates. The used evaluation metrics are as follows:

\begin{itemize}
  \item \textit{Recall@$K$}: It measures the proportion of correctly predicted relevant items in the top-$K$ recommendation list:
  \begin{equation}
    \text{Recall@}K = \frac{1}{|\mathcal{U}|} \sum_{u \in \mathcal{U}} \mathbb{I}(r_{u} < K),
  \end{equation}
  where $\mathcal{U}$ is the set of users in the test set, and $r_u$ denotes the rank position of the ground-truth item $i_u$ in the top-$K$ recommendation list for user $u$.
  
  \item \textit{NDCG@$K$}: It evaluates the ranking quality of the top-$K$ recommended items by considering both item relevance and order:
  \begin{equation}
    \text{DCG}_u@K = \sum_{j=1}^{K} \frac{2^{y_{u,j}} - 1}{\log_2(j + 1)},
  \end{equation}
  \begin{equation}
    \text{iDCG}_u@K = \sum_{j=1}^{K} \frac{2^{y^*_{u,j}} - 1}{\log_2(j + 1)},
  \end{equation}
  \begin{equation}
    \text{NDCG}_u@K = \frac{\text{DCG}_u@K}{\text{iDCG}_u@K},
  \end{equation}
  \begin{equation}
    \text{NDCG@}K = \frac{1}{|\mathcal{U}|} \sum_{u=1}^{|\mathcal{U}|} \text{NDCG}_u@K,
  \end{equation}
  where $y_{u,j}$ denotes the relevance of the $j$-th recommended item for user $u$, and $y^*_{u,j}$ is the relevance in the ideal ranking. In binary relevance settings, $y_{u,j}=1$ if the $j$-th item is the ground-truth item, and $0$ otherwise. The final NDCG@$K$ is the average over all users.
\end{itemize}

Following the standard evaluation protocol \cite{kang2018self}, we use the leave-one-out strategy for evaluation. For each item sequence, the last item is used for testing, the item before the last is used for validation, and the rest is used for training. During training, we limit the number of items in a user’s history to 20.

\subsubsection{Implementation Details} 
In our experiments, we set the number of encoder layers to 1 and the number of decoder layers to 4. Following previous work \cite{feng2022recommender, wang2024eager}, we adopt DIN as our sequential recommendation model and use pretrained Llama2-7b as our semantic encoder, with the hidden size set to 128 as reported in \cite{rajput2023recommender, wang2024eager}. The number of clusters in the hierarchical k-means is set to 256. To train our model, we use the Adam optimizer with a learning rate of 0.001 and apply a warmup strategy for stable training. The batch size is set to 256. Notably, UNGER demonstrates robustness to the hyperparameters of the CKA and IKD tasks, attributed to their fast convergence. Consequently, the loss coefficients $\alpha$ and $\beta$ are both set to 1. Each experiment is conducted five times and the average score is reported. All model parameters and settings are shown in Table \ref{tab:model_params}.

\begin{table}[htbp]
\centering
\caption{Model hyperparameters and settings.}  
{
\begin{tabular}{@{}ll@{}}
\toprule
\textbf{Parameter} & \textbf{Value} \\ 
\midrule
Embedding Dimension & 96 \\
Model Layers    & 4 \\
Hidden Size         & 256 \\
Heads    & 6 \\
Num of Clusters      & 256  \\
Cluster Depth & 4 \\
Learning Rate       & 1e-3 \\
Optimizer           & Adam \\
Semantic Encoder    & Llama2-7b \\
Collaborative Encoder  & DIN \\
Batch Size          & 256 \\
Training Steps     & 20000 \\
Dropout Rate        & 0 \\
Beam Width          & 100 \\
Activation Function & ReLU \\
Weight Decay        & 1e-7 \\
Warmup Steps        & 2000 \\
Warmup Initial Learning Rate    & 1e-7 \\
Random Seed       & 2024 \\
$\alpha$     & 1.0 \\
$\beta$        & 1.0 \\
$\tau$       & 1.0 \\
\bottomrule
\end{tabular}
}
\label{tab:model_params}
\end{table}

\begin{table}[htbp]
\centering
\caption{The information about practical resources required.}  
\begin{tabular}{ll}
\toprule
\textbf{Information} & \textbf{Value} \\
\midrule
Device & $\text{NVIDIA RTX 3090 GPU (24GB)} \times 1$ \\
Training Time & less than 1 GPU hour \\
Inference Latency (ms) & 19.5 \\
Model Parameter & 10249988 \\
\bottomrule
\end{tabular}
\label{tab:training resource}
\end{table}

\subsection{Performance Comparison (RQ1)}
\subsubsection{Baselines}We compare UNGER with several representative related baselines, encompassing classical sequential modeling approaches as well as the latest emergent generative techniques.

(1) For \textit{classical sequential} methods, we have:
\begin{itemize}
    \item \textbf{GRU4REC} \cite{hidasi2015session} is an RNN-based model that utilizes Gated Recurrent Units (GRUs) to model user click sequences.
    \item \textbf{Caser} \cite{tang2018personalized} is a CNN-based method capturing high-order Markov Chains by modeling user behaviors through both horizontal and vertical convolutional operations.
    \item \textbf{SASRec} \cite{kang2018self} is a self-attention-based sequential recommendation model that utilizes a unidirectional Transformer encoder with a multi-head attention mechanism to effectively model user behavior and predict the next item in a sequence.
    \item \textbf{BERT4Rec} \cite{sun2019bert4rec} adopts a Transformer model with the bidirectional self-attention mechanism and uses a Cloze objective loss for the modeling of item sequences.
    \item \textbf{HGN} \cite{ma2019hierarchical} adopts hierarchical gating networks to capture long-term and short-term user interests.
    \item \textbf{FDSA} \cite{zhang2019feature} leverages self-attention networks to separately model item-level and feature-level sequences, emphasizing the transformation patterns between item features and utilizing a feature-level self-attention block to capture feature transition dynamics.
    \item \textbf{S$^3$-Rec} \cite{zhou2020s3} employs mutual information maximization to pre-train a bi-directional Transformer for sequential recommendation, enhancing the model's ability to learn correlations between items and their attributes through self-supervised tasks.
\end{itemize}

(2) For \textit{generative} methods, we have:
\begin{itemize}
    \item \textbf{RecForest} \cite{feng2022recommender} jointly learns latent embeddings and indices through multiple \(K\)-ary trees, utilizing hierarchical balanced clustering and a transformer-based encoder-decoder routing network to enhance accuracy and memory efficiency in identifying top-n items.
    \item \textbf{TIGER} \cite{rajput2023recommender} utilizes a pre-trained T5 encoder to learn semantic identifiers for items, autoregressively decodes target candidates using these identifiers, and incorporates RQ-VAE quantization to build generalized item identifiers without relying on sequential order.
    \item \textbf{ColaRec} \cite{wang2024enhanced} integrates user-item interactions and content data within an end-to-end framework, leveraging pretrained collaborative identifiers, an item indexing task, and contrastive loss to align semantic and collaborative spaces for enhanced recommendation performance.
    \item \textbf{EAGER} \cite{wang2024eager} integrates behavioral and semantic information through a two-stream architecture with shared encoding, separate decoding, and strategies for contrastive and semantic-guided learning to enhance collaborative information utilization.
\end{itemize}

\subsubsection{Results} Table \ref{tab:overall performance} reports the overall performance of three datasets. The results for all baselines without the superscript $^*$ are taken from the publicly accessible results \cite{zhou2020s3, wang2024eager}. For missing statistics, we reimplement the baseline and report our experimental results. From the results, we have the following observations:
\begin{itemize}
    \item \textbf{UNGER consistently achieves superior performance compared to base models across multiple datasets.} 
    Specifically, on the Beauty dataset, our proposed model UNGER outperforms the second-best baseline, EAGER, by a substantial margin, achieving improvements of 14.68\% in \textit{Recall@20} and 7.85\% in \textit{NDCG@20}. These results underscore the effectiveness of our approach in real-world recommendation settings. We attribute this significant performance gain to UNGER's novel ability to effectively integrate the complementary strengths of collaborative filtering signals and semantic content representations through a unified discrete code. By employing a compact representation space, UNGER not only enhances model efficiency but also reduces redundancy, achieving better performance with approximately half the number of total parameters compared to EAGER. This makes UNGER not only more accurate but also more lightweight and scalable.

    \item \textbf{UNGER surpasses prior generative recommendation models on most datasets.} 
    In head-to-head comparisons with previous generative baselines, UNGER exhibits a consistent advantage in performance metrics. This superiority stems from our model’s two-stage multi-task joint optimization framework, which is specifically designed to capture and unify collaborative and semantic knowledge within a shared code space. This framework enables UNGER to learn more coherent and expressive item representations by reinforcing the mutual information between collaborative signals and semantic embeddings. The unified code serves as a bridge that tightly couples these two modalities, thereby allowing the model to make more informed and nuanced recommendations. These improvements affirm the validity of our design and emphasize the importance of unified code representations in generative recommendation models.

    \item \textbf{Generative models generally outperform classical recommendation models in most cases.} 
    Our results reveals a clear trend: generative models tend to outperform traditional approaches such as sequential models based purely on item IDs. A key differentiator lies in how items are represented. Traditional methods often treat item IDs as atomic, non-informative symbols, lacking any structural or contextual information. In contrast, generative models, especially those based on Transformer decoders, leverage discrete hierarchical codes that encapsulate rich prior knowledge about item semantics and relationships. This structured representation aligns naturally with the token-by-token generation process of Transformers, enabling the model to better capture sequential dependencies and contextual relevance. As a result, generative models not only deliver better accuracy but also provide a more interpretable and principled framework for recommendation generation.
\end{itemize}

\begin{figure}[t]
  \centering
  \includegraphics[width=\linewidth]{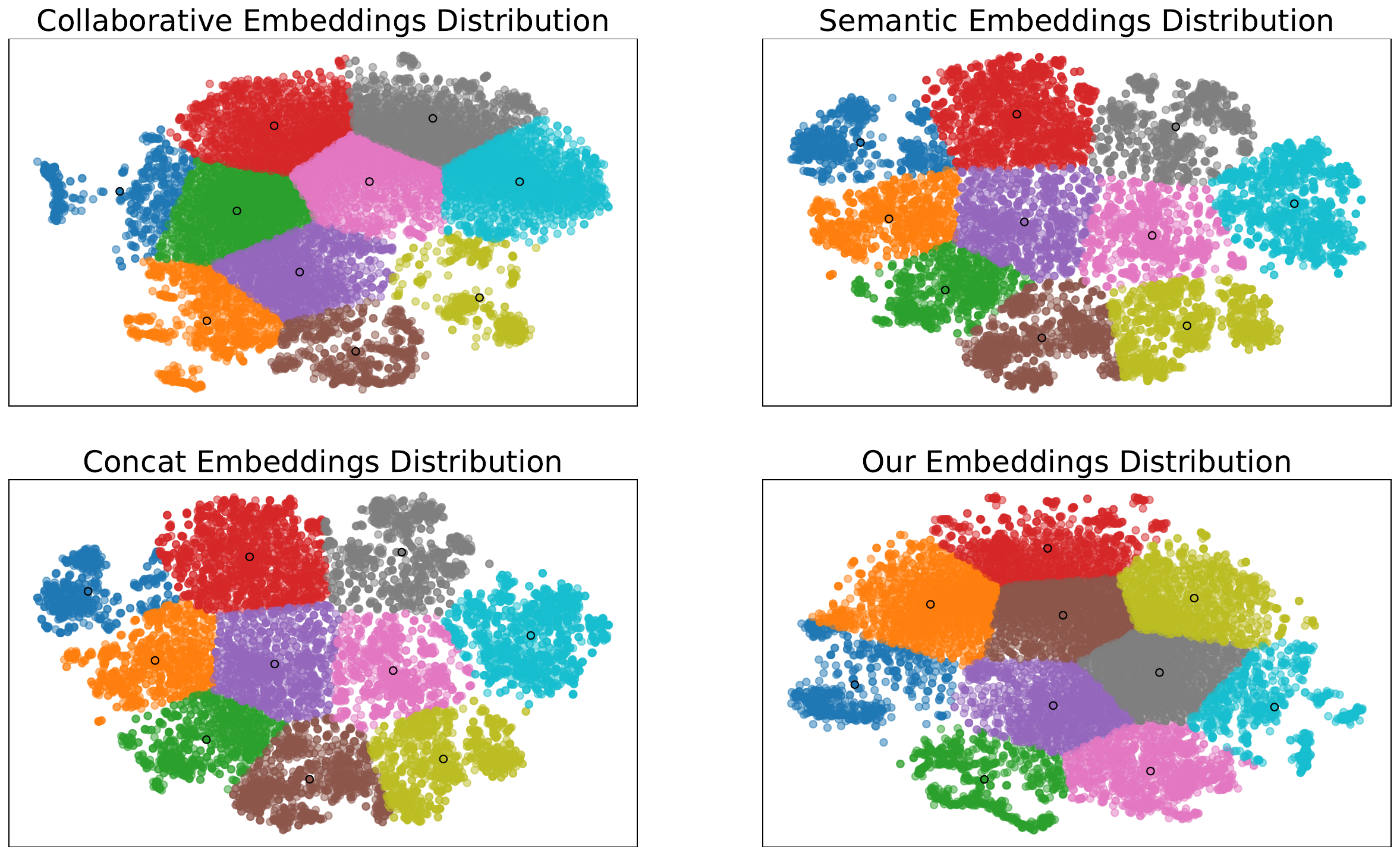}
  \caption{T-SNE visualization of different embeddings distribution after applying K-means clustering (clusters=10).}
  \Description{}
  \label{fig:distribution}
\end{figure}

\begin{table}[ht]
  \caption{Proportional similarity of the semantic modality and collaborative modality to the final representation across different approaches.}
  \label{tab:relative influence}
  \resizebox{0.6\linewidth}{!}{
  \begin{tabular}{ccc}
    \toprule
    Approach & Semantic modality & Collaborative modality\\
    \midrule
    Concat & 97.33\% & 2.67\% \\
    Ours & 59.89\% & 40.11\% \\
    \bottomrule
\end{tabular}
}
\end{table}

\begin{table}[htbp]
    \centering
    \caption{Performance of different methods.}
    \label{tab:fusion methods}
    \resizebox{0.7\linewidth}{!}{
    \begin{tabular}{lcccc}
    \toprule
    Dataset & \multicolumn{4}{c}{Beauty} \\
    \cmidrule(lr){2-5}
    Metric & Recall@10 & Recall@20 & NDCG@10 & NDCG@20 \\
    \midrule
    Semantic Only     & 0.0791 & 0.1077 & 0.0462 & 0.0535 \\
    Collaborative Only & 0.0744 & 0.0997 & 0.0469 & 0.0532 \\
    Concat            & 0.0759 & 0.1071 & 0.0452 & 0.0530 \\
    Ours     & \textbf{0.0939} & \textbf{0.1289} & \textbf{0.0559} & \textbf{0.0646} \\
    \bottomrule
    \end{tabular}
    }
\end{table}

\subsection{Semantic Domination Issue (RQ2)}
To investigate whether our method addresses the semantic dominance issue, we conduct experiments on the Amazon Beauty dataset. We first extract the semantic embeddings of the items, then use only the item IDs from historical interaction data and apply the classical DIN model to obtain the collaborative embeddings. Following the most widely used vanilla approach of fusing knowledge from different modalities \cite{peng2022balanced}, concatenation, we concatenate the semantic embeddings with the collaborative embeddings. Notably, to mitigate the potential issues arising from mismatches in both dimensionality and value distribution between the two modalities, we first apply PCA to reduce the dimension of the semantic embeddings to match that of the collaborative embeddings. Then, we apply normalization to both embeddings before concatenating them to obtain the final concatenation embeddings. In addition, we obtain the integrated embeddings of items using our proposed method. We use T-SNE to visualize the distribution of these different embeddings. To more clearly illustrate the differences in their distributions, we apply K-means clustering algorithm to these embeddings, using different colors to represent data points from each cluster, as shown in Figure \ref{fig:distribution}.

Furthermore, to quantitatively measure the relative similarity of semantic embeddings and collaborative embeddings on the final embeddings using different methods, we use KL divergence to assess the similarity of their distribution. The relative similarity of each modality is then calculated based on the KL divergence values. The formula for above process is as follows:
\begin{equation}
\text{Similarity}_{\text{semantic}} = 1-\frac{{\text{KL}}(S \parallel E)}{{\text{KL}}(S \parallel E) + {\text{KL}}(C \parallel E)}
\end{equation}

\begin{equation}
\text{Similarity}_{\text{collaborative}} = 1-\frac{{\text{KL}}(C \parallel E)}{{\text{KL}}(S \parallel E) + {\text{KL}}(C \parallel E)}
\end{equation}
\noindent where \( E \) represents the final embeddings using different methods, \( S \) denotes the semantic embeddings, and \( C \) refers to the collaborative embeddings.

As shown in Table \ref{tab:relative influence}, the concatenation method yields notably different similarity proportions between the semantic and collaborative modalities. Figure \ref{fig:distribution} further reveals that the resulting fused embeddings share a distributional pattern closely aligned with the original semantic embeddings. We refer to this phenomenon as the semantic dominance problem, which we define as a distributional imbalance where semantic modality disproportionately influences the final representation, thereby diminishing the contribution of the collaborative modality. While the two modalities are embedded in spaces of equal dimensionality and similar value ranges, their intrinsic data characteristics (e.g., semantic embeddings often being pretrained on large-scale textual data with smooth distributional properties) can lead to asymmetric influence when directly concatenated. Concatenation method tends to overly emphasize semantic knowledge while greatly ignoring collaborative knowledge, which poses a substantial obstacle to downstream recommendation effectiveness. In contrast, our proposed method effectively addresses this issue. It not only learns embeddings from both collaborative and semantic modalities in a more balanced manner at the distributional level but also achieves a more proportional similarity from each modality in quantitative evaluations. This prevents the semantic embeddings from dominating the final representation. As illustrated in Table \ref{tab:fusion methods}, our method significantly outperforms both single-modality approaches and the concatenation-based fusion method, achieving the best results across all evaluation metrics. Notably, the concatenation method even performs worse than the semantic-only approach, indicating a negative impact on performance. This further highlights the severity of the semantic dominance problem and underscores the advantages of our proposed approach. We attribute these improvements to the successful integration of collaborative and semantic knowledge, which, unlike the concatenation method, better captures the complementary nature of the two different knowledge, enabling a more balanced and effective fusion of information from both modalities.

\begin{table}[t]
    \centering
    \caption{Performance comparison of different variants.}
    \label{tab:ablation study}
    \renewcommand{\arraystretch}{1.2}
    \begin{tabular}{ccccc}
        \toprule
        \textbf{Dataset} & \textbf{Metric} & \textbf{\textbackslash} & \textbf{CKA} & \textbf{CKA + IKD (Ours)} \\
        \midrule
        \multirow{4}{*}{Beauty}
        & Recall@10   & 0.0759 & 0.0827 & 0.0939 \\
        & Recall@20   & 0.1071 & 0.1122 & 0.1289 \\
        & NDCG@10     & 0.0452 & 0.0509 & 0.0559 \\
        & NDCG@20     & 0.0530 & 0.0583 & 0.0646 \\
        \bottomrule
    \end{tabular}
\end{table}

\subsection{Ablation Study (RQ3)}



To better understand the individual contributions of the components in our proposed model, UNGER, we conducted a comprehensive ablation study. Specifically, we incrementally removed two key auxiliary tasks, intra-modality knowledge distillation (IKD) task and cross-modality knowledge alignment (CKA) task to construct several ablated versions of the model. The corresponding performance of each variant is reported in Table~\ref{tab:ablation study}.

\begin{itemize}
    \item Impact of Removing IKD and CKA. We observe that removing either the CKA or IKD task leads to a noticeable degradation in performance. When both tasks are excluded, the model reduces to a basic architecture where semantic and collaborative embeddings are merely concatenated. This baseline yields the poorest performance among all variants. These results strongly suggest that both IKD and CKA play indispensable roles in the success of UNGER.
    
    \item Effectiveness of the CKA Task. When only the CKA task is retained, the model already exhibits significant performance gains over the baseline. This demonstrates that aligning knowledge across collaborative and semantic modalities allows the model to capture richer and more holistic item representations. The CKA task serves as a bridge between modalities, allowing the model to harness their complementary information effectively.
    
    \item Synergistic Effect of CKA and IKD. The best performance is achieved when both the CKA and IKD tasks are jointly incorporated. In this setting, CKA establishes strong cross-modal connections while IKD reinforces intra-modal coherence, enabling the model to fully exploit the complementary strengths of collaborative and semantic knowledge. This synergy facilitates the learning of powerful, unified item codes that significantly enhance recommendation quality.
\end{itemize}

In conclusion, the results of the ablation study clearly demonstrate that each component of UNGER contributes meaningfully to its overall performance. The IKD and CKA tasks are not only effective in isolation but also highly complementary. Their integration is essential to achieving state-of-the-art results in generative recommendation, validating the effectiveness and superiority of our method.


\begin{figure}[t]
  \centering
  \includegraphics[width=\linewidth]{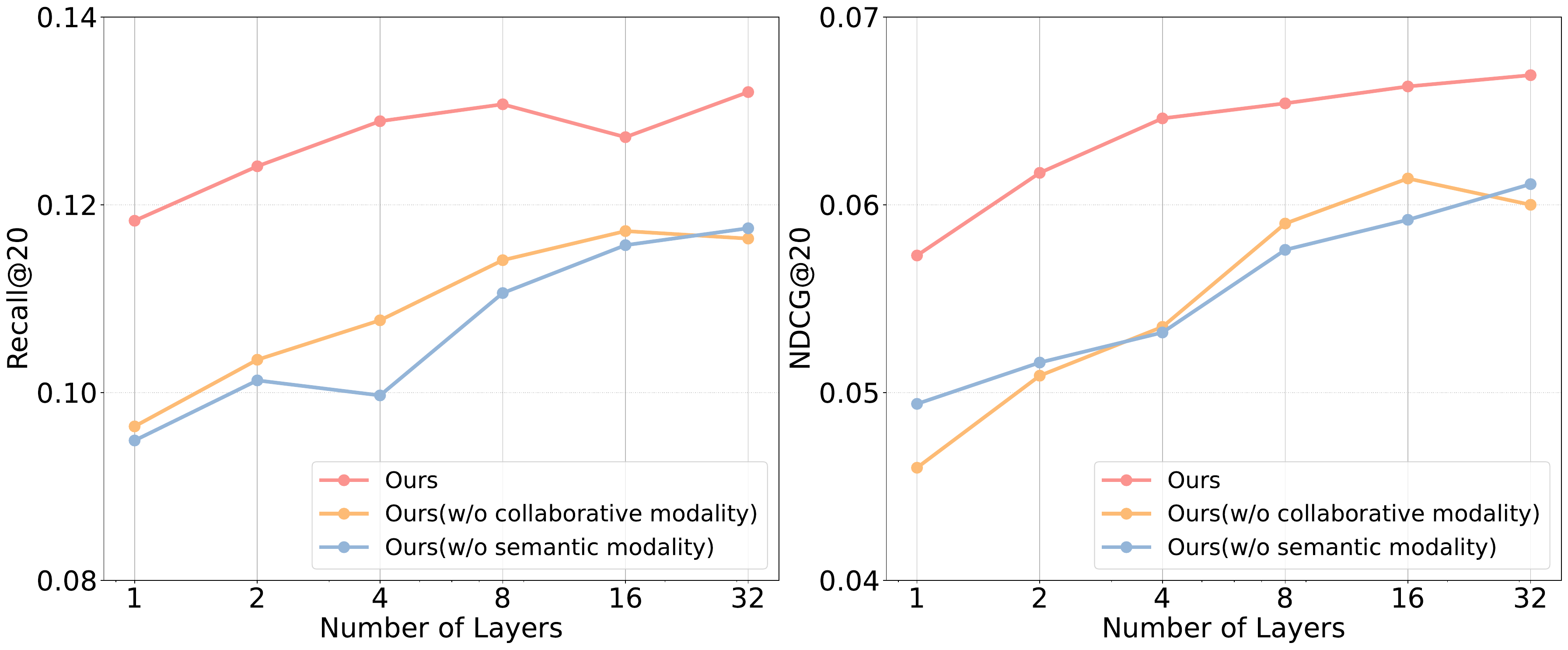}
  \caption{Impact of the number of model layers on the Beauty dataset. The number of layers varies from 1 to 32.}
  \Description{}
  \label{fig:model layer}
\end{figure}

\subsection{Scaling Law Study (RQ4)}
Generative models have gained considerable attention due to their potential to exhibit scaling law properties. However, research exploring this phenomenon in the context of generative recommendation remains limited. Although a recent study, HSTU \cite{zhai2024actions}, has investigated the scaling law characteristics of generative recommendation, it is restricted to collaborative signals derived purely from item IDs and only examines the impact of model depth. This narrow focus limits a more comprehensive understanding of scaling behavior. In contrast, to the best of our knowledge, no prior studies have explored scaling laws in models that jointly leverage both semantic and collaborative information. In this work, we make an initial and systematic attempt to explore the scaling behavior of generative recommendation models, examining how model performance evolves with respect to model depth, model width, and data volume.
\subsubsection{Model Depth}
To investigate the effect of model depth on generative recommendation performance, we vary the number of layers from 1 to 32 and evaluate the models on the \textit{Beauty} dataset. As illustrated in Figure \ref{fig:model layer}, the overall trend indicates consistent performance improvement as the model layers increases. Both Recall@20 and NDCG@20 show steady gains, particularly within the range of 1 to 8 layers, demonstrating the increasing capacity of deeper models to capture complex user-item interaction patterns.

We attribute the superior performance of our method to the unified code structure, which jointly encodes collaborative and semantic information into a shared representation space. This unified design enables the model to leverage both low-level interaction signals and high-level semantic attributes, thus offering richer and more diverse learning signals at every layer. In addition, the unified code structure mitigates redundancy between modalities and facilitates mutual regularization, allowing the model to disentangle meaningful patterns more effectively as it scales.

This synergy becomes increasingly beneficial with greater model depth. Deeper layers are able to refine and abstract over both types of information in a complementary manner, enhancing the model’s ability to generalize without overfitting to a single modality. Consequently, our full model consistently outperforms its ablated variants across all depth settings, and the performance gap widens as depth increases, highlighting the advantage of integrating both semantic and collaborative modalities in a unified framework.

\begin{figure}[t]
  \centering
  \includegraphics[width=0.7\linewidth]{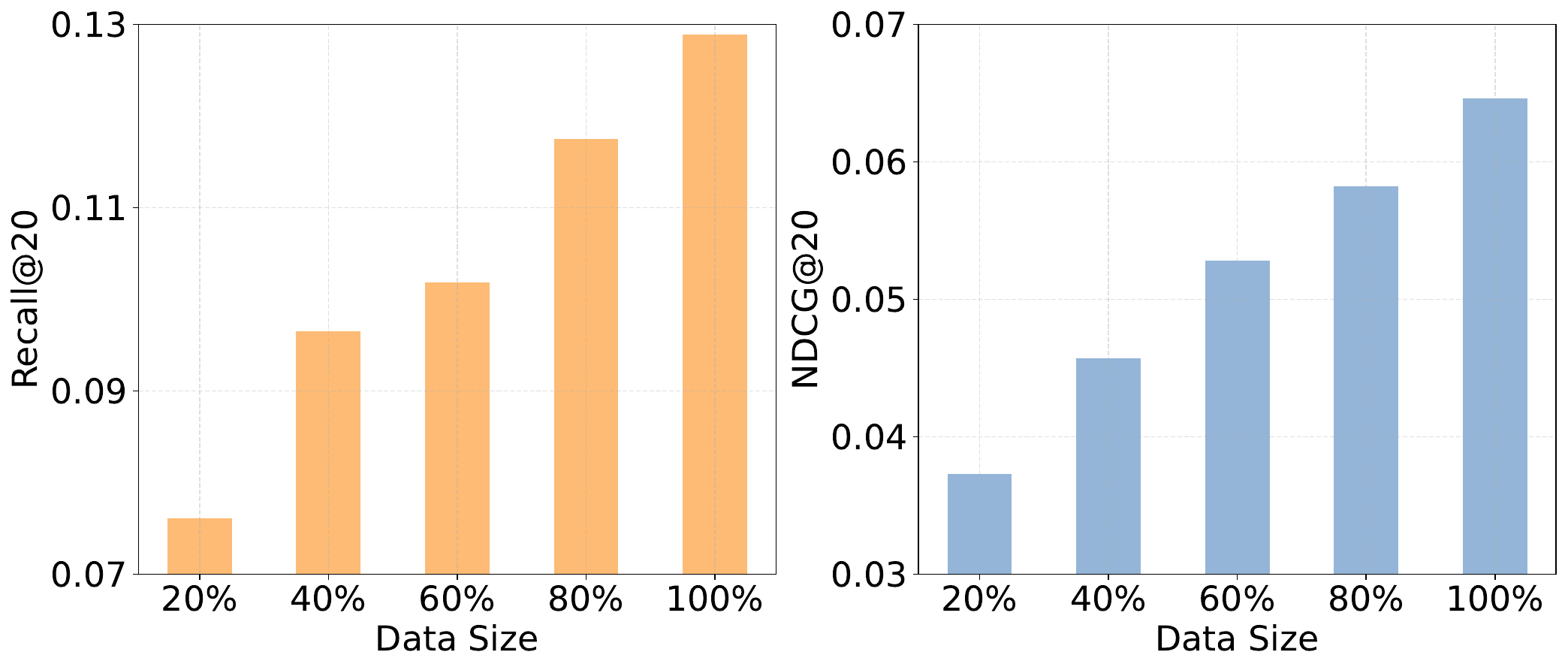}
  \caption{Impact of data volume on the Beauty dataset.}
  \Description{}
  \label{fig:data size}
\end{figure}

\begin{figure}[t]
  \centering
  \includegraphics[width=0.7\linewidth]{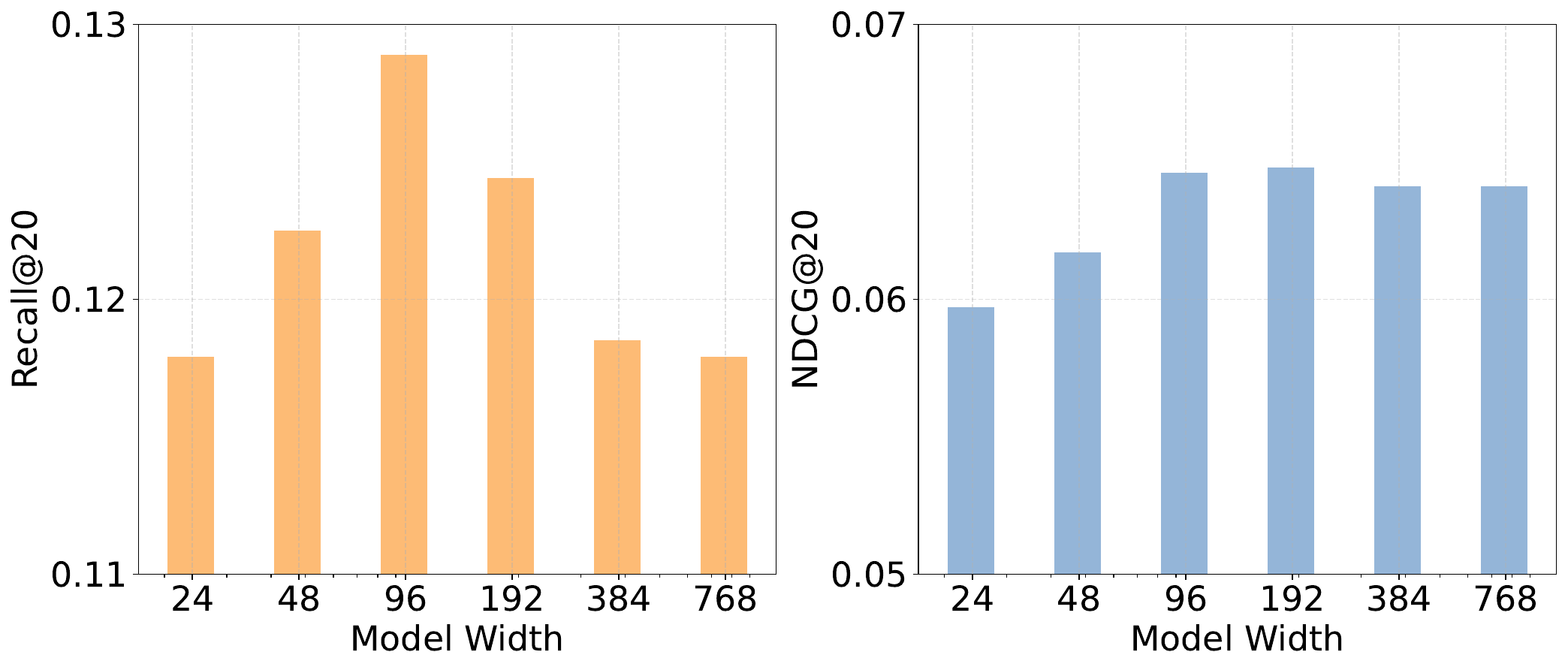}
  \caption{Impact of model width on the Beauty dataset. The number of width varies from 24 to 768.}
  \Description{}
  \label{fig:model width}
\end{figure}

\subsubsection{Data Size}
We further examine how data volume influences the performance of generative recommendation by training models on subsets of the \textit{Beauty} dataset, ranging from 20\% to 100\% of the full data. As shown in Figure \ref{fig:data size}, both Recall@20 and NDCG@20 consistently improve with increasing data size, suggesting that the model effectively benefits from larger training corpora.

This upward trend underscores the data-efficient nature of our unified code structure, which jointly leverages collaborative and semantic modalities. By learning from both interaction histories and item semantics, the model is able to extract more meaningful patterns as the data scale increases. In particular, Recall@20 experiences a significant gain when the data volume increases from 20\% to 40\%, suggesting that the model can rapidly leverage additional interaction signals to enhance its effectiveness. Beyond this point, model performance continues to improve steadily with more data, highlighting the scalability of our approach. The consistent upward trend across all data volumes demonstrates the importance of training data scale in enhancing the capacity of generative recommendation models.

\subsubsection{Model Width}
Besides, We also investigate how model width influences the performance of generative recommendation by varying the embedding dimension from 24 to 768. As presented in Figure \ref{fig:model width}, both Recall@20 and NDCG@20 exhibit an initial upward trend, followed by a performance decline as the width continues to increase.

The model achieves the best performance when the width is set to 96, indicating that a moderate level of width provides a good balance between model capacity and generalization. Increasing the width beyond this point does not lead to further improvement and even results in decreased performance, likely due to over-parameterization and reduced training stability under limited data conditions.

These results highlight the importance of selecting an appropriate width to fully utilize the benefits of our unified code structure. While wider models offer higher representational capacity, overly large widths may introduce optimization difficulties and lead to suboptimal generalization. Therefore, a carefully tuned width is essential for generative recommendation.

\begin{figure}[t]
    \centering
    \includegraphics[width=0.6\linewidth]{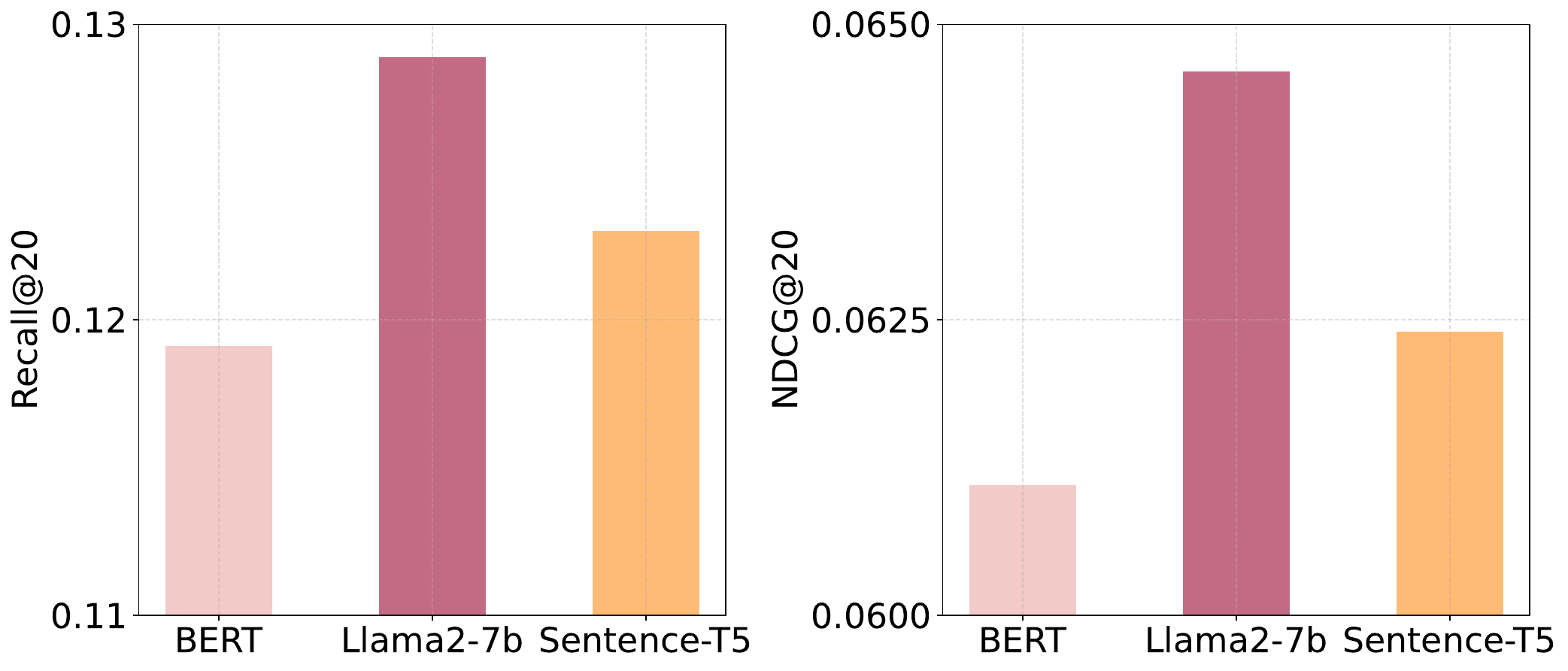}
    \caption{Analysis of different pretrained semantic encoders.}
    \label{fig:semantic_encoder}
\end{figure}

\begin{figure}[t]
    \centering
    \includegraphics[width=0.6\linewidth]{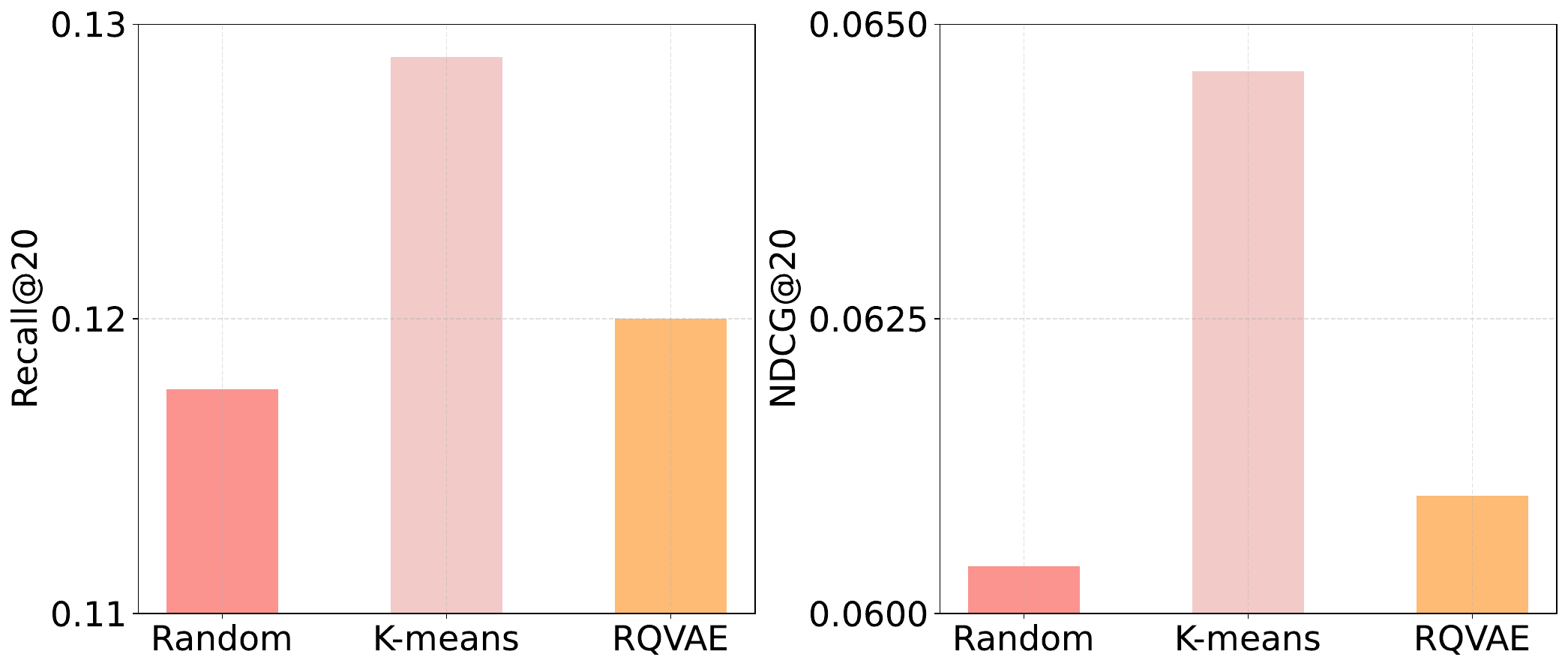}
    \caption{Analysis of different quantization methods.}
    \label{fig:quantization_methods}
\end{figure}

\begin{figure}[t]
    \centering
    \includegraphics[width=0.6\linewidth]{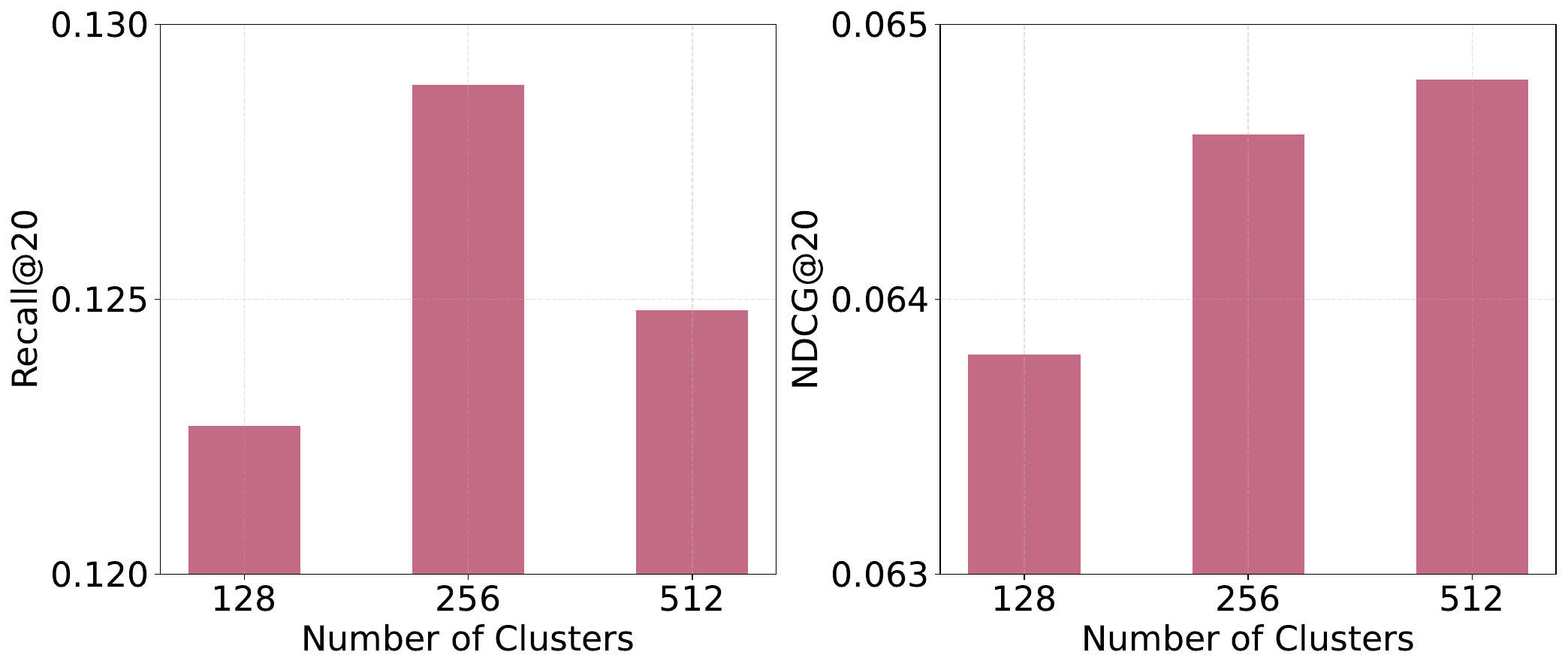}
    \caption{Impact of the number of clusters with $K$ varying from 128 to 512.}
    \label{fig:number of cluster}
\end{figure}

\subsection{Hyper-Parameter Analysis (RQ4)}
\subsubsection{Choice of different semantic encoders}
To investigate the influence of different pretrained semantic encoders on recommendation performance, we conduct a comprehensive comparative analysis on the Beauty dataset. Specifically, we explore three representative language models for extracting item-level textual representations: (1) \textbf{BERT} \cite{devlin2019bert}, a bidirectional transformer trained with a masked language modeling objective, commonly used for general-purpose language understanding tasks; (2) \textbf{Sentence-T5} \cite{ni2021sentence}, an encoder-decoder model optimized for semantic similarity and text matching, particularly effective on short-text tasks; and (3) \textbf{Llama2-7b} \cite{touvron2023llama}, a large-scale decoder-only model trained on a diverse and extensive corpus, capable of capturing rich contextual and domain-specific semantics. These language models serve as plug-and-play components within our framework, allowing us to isolate and compare their impact on downstream recommendation quality under a unified training and evaluation protocol.

As presented in Figure~\ref{fig:semantic_encoder}, the results reveal the following insights: (1) Llama2-7b achieved the best performance, likely due to its substantially larger and more diverse training corpus, which covers a wide range of domains including news articles, encyclopedic content, social media posts, and source code. This extensive and heterogeneous data exposure enables Llama2-7b to learn richer contextual semantic relationships and finer-grained concept representations. In contrast, Sentence-T5, although specifically optimized for short-text semantic matching, is primarily trained on translation and summarization tasks. As a result, it struggles to fully capture the diverse, domain-specific, and stylistically nuanced nature of e-commerce product descriptions, which often involve specialized terminology and dynamic fashion-related expressions. BERT, while known for its stability on general-purpose language understanding tasks, employs a bidirectional masked language modeling objective that is not explicitly designed for semantic generation. Consequently, its embeddings are less effective in distinguishing polysemous words, capturing long-tail attributes, and modeling fine product-level distinctions. These limitations reduce the effectiveness of downstream residual quantization and semantic transfer in the generative recommendation stage, leading to suboptimal performance. (2) Our method is inherently plug-and-play, meaning that it allows the semantic encoder to be replaced without altering the residual quantization mechanism or the generative decoder. This design makes it easy to integrate the latest advancements in language modeling, such as DeepSeek-R1 \cite{guo2025deepseek}, GPT-4o \cite{hurst2024gpt}, or domain-specific encoders trained on vertical data like fashion or electronics. Furthermore, our method can be extended to incorporate multimodal encoders that combine textual and visual inputs, such as CLIP-like models \cite{radford2021learning, liu2023visual, li2022blip}, as long as the output dimensions are compatible. This modular and extensible architecture ensures that our framework can continually benefit from progress in pretrained representation learning, leading to consistent improvements in recommendation performance across domains.

\subsubsection{Choice of different quantization methods.}

To evaluate the impact of different quantization strategies on model performance, we conduct a comprehensive comparative analysis on the Beauty dataset. Specifically, we consider three distinct methods for discretizing continuous item representations: (1) \textbf{Random Assignment}, where items are randomly mapped to discrete codes without considering semantic or structural relationships; (2) \textbf{RQVAE} \cite{lee2022autoregressive}, which performs multi-level residual vector quantization through deep learning to capture high-dimensional semantics; and (3) \textbf{Hierarchical K-means}, a tree-structured clustering method that recursively partitions the embedding space to preserve hierarchical relations among items.

As illustrated in Figure~\ref{fig:quantization_methods}, the hierarchical K-means approach significantly outperforms the other two methods across various evaluation metrics. This result highlights the importance of preserving structural information during the quantization process. In particular, hierarchical K-means encodes items such that those sharing common prefix paths in the hierarchical tree tend to be semantically or functionally similar. This structure enables the model to exploit hierarchical proximity during recommendation, improving the likelihood of retrieving items relevant to user preferences. 

In contrast, the random assignment method fails to provide any meaningful structure to the discretized space, resulting in a noticeable drop in performance. The codes in this method do not reflect any semantic similarity, thereby limiting the model's ability to generalize based on code proximity. While the RQVAE method is theoretically capable of modeling complex item distributions through its residual quantization mechanism, in practice we observe that it frequently suffers from \textit{codebook collisions}. This refers to the phenomenon where multiple semantically distinct items are assigned to the same code due to limited codebook diversity or suboptimal training dynamics. Such collisions reduce the representational capacity of the quantized space and negatively impact recommendation performance. To address this issue, we adopt a mitigation strategy inspired by prior work~\cite{rajput2023recommender}, wherein a unique identifier is appended to the learned code to enforce item-level uniqueness. Although this heuristic alleviates the collision problem, it compromises the learned hierarchical structure, as the appended identifier does not reflect semantic similarity. Consequently, this workaround introduces noise into the code space and may lead to suboptimal downstream performance. 

Overall, these observations collectively underscore the advantages of hierarchical quantization methods in capturing and preserving the structural and semantic relationships among items, which are crucial for effective recommendation.

\subsubsection{Number of Clusters.} 

We conduct an empirical investigation into how the number of clusters \(K\) in the hierarchical K-means algorithm influences the overall performance of our method on the Beauty dataset. In particular, we experiment with three different cluster sizes: \(K = 128\), \(K = 256\), and \(K = 512\). The results are summarized in Figure~\ref{fig:number of cluster}.

Our analysis reveals that increasing \(K\) from 128 to 256 leads to a consistent improvement in both Recall@20 and NDCG@20 metrics. This improvement is likely due to the enhanced representational capacity of the codebook as the number of clusters increases. When \(K\) is relatively small (e.g., 128), the limited number of cluster centroids restricts the ability of the model to assign unique or highly discriminative codes to individual items. As a result, semantically or behaviorally dissimilar items may be forced to share similar codes, thereby reducing the granularity and accuracy of the learned item representations.

However, when we further increase \(K\) from 256 to 512, we observe a more nuanced outcome. Specifically, the NDCG@20 metric continues to improve, indicating that the ranking quality of the top predicted items benefits from the more expressive codebook. This suggests that the model becomes better at placing the truly relevant items at higher ranks in the recommendation list. On the other hand, the Recall@20 metric slightly declines in this setting. We hypothesize that this is due to the expanded search space introduced by a larger \(K\), which increases the decoding complexity and may lead to suboptimal retrieval results during inference. In other words, while the model has more representational freedom, the difficulty of accurately decoding from a much larger codebook may offset some of the gains.

Anyway, these findings collectively suggest that in generative recommendation systems, the number of clusters \(K\) used in hierarchical K-means should be chosen carefully to strike a balance between representational richness and computational efficiency. A value of \(K\) that is too small can constrain the model's ability to encode item uniqueness, while an excessively large \(K\) can introduce challenges in decoding and generalization. Selecting a cluster size proportional to the overall number of items in the dataset appears to be a practical heuristic for achieving this balance.

\begin{figure}[t]
    \centering
    \includegraphics[width=0.6\linewidth]{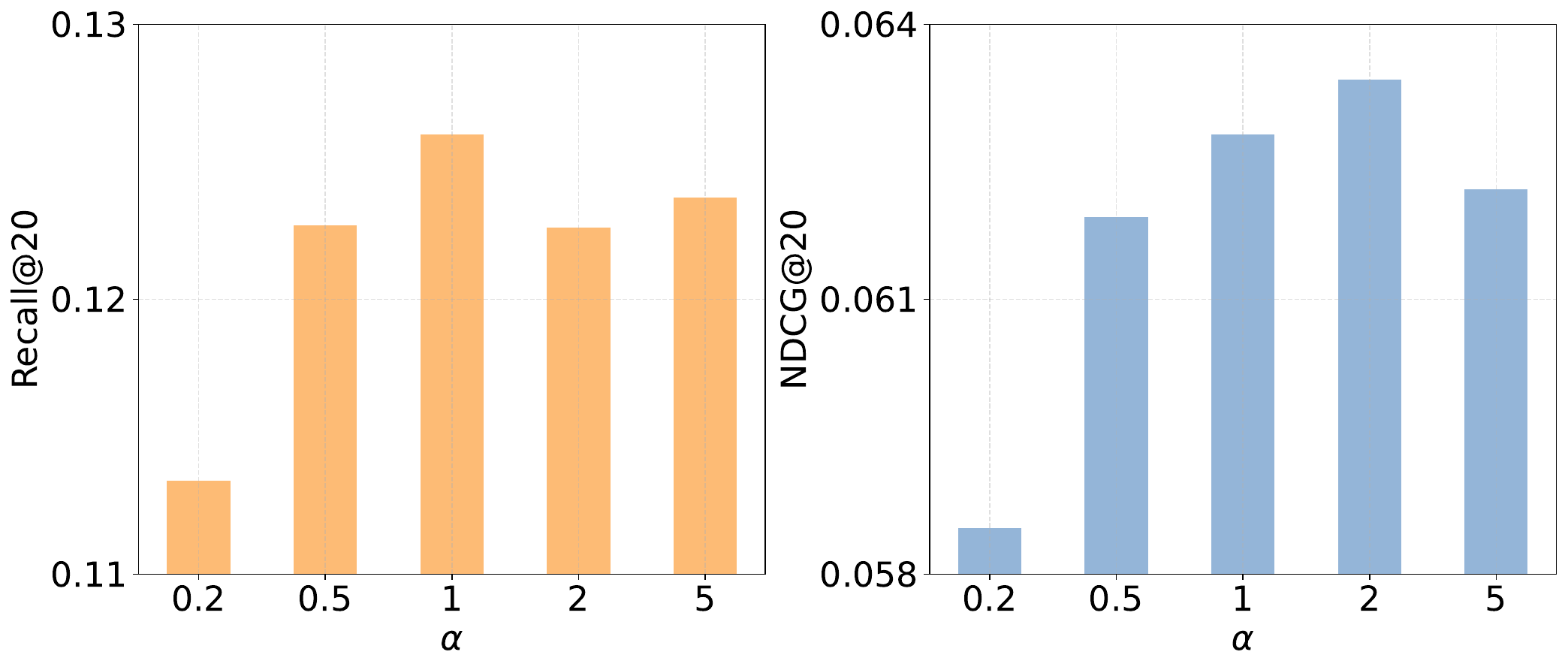}
    \caption{Sensitivity of hyper-parameter $\alpha$}
    \label{fig:sensitivity of alpha}
\end{figure}

\begin{figure}[t]
    \centering
    \includegraphics[width=0.6\linewidth]{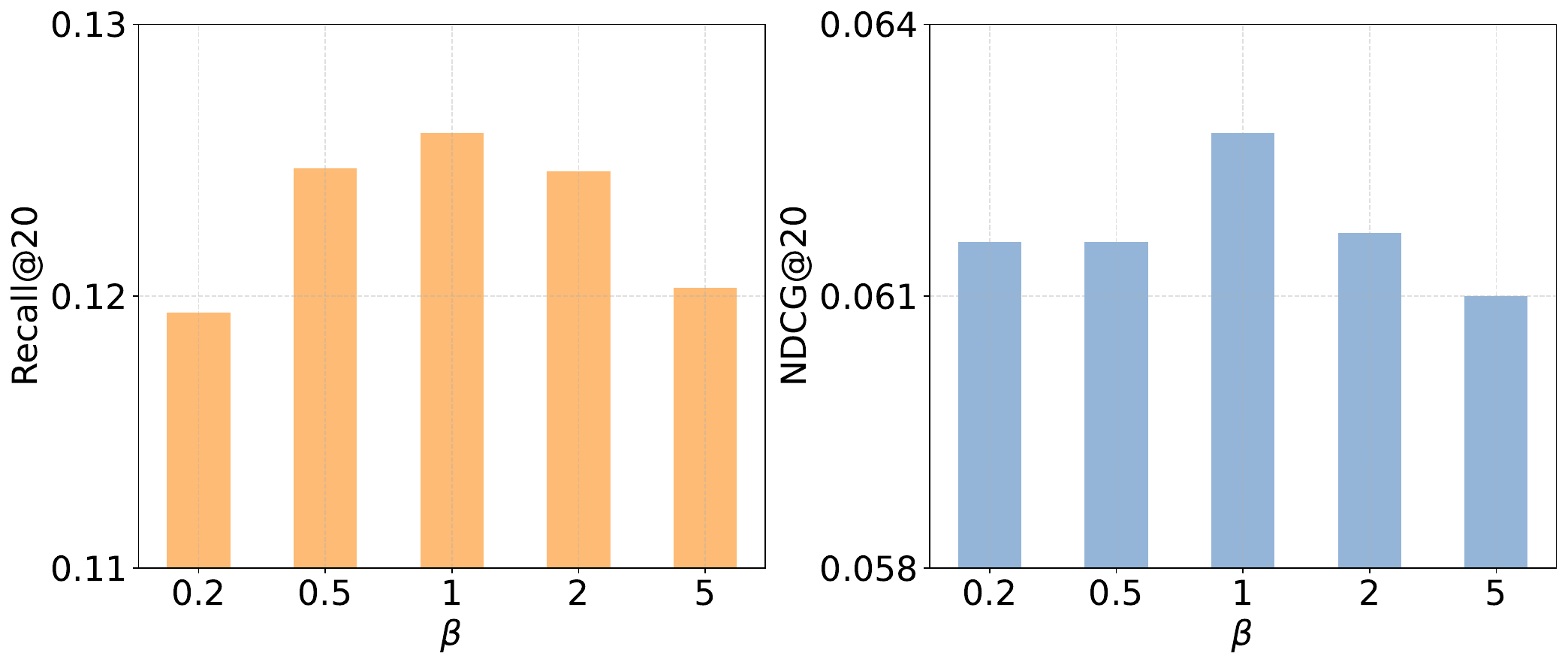}
    \caption{Sensitivity of hyper-parameter $\beta$}
    \label{fig:sensitivity of beta}
\end{figure}

\subsubsection{Sensitivity of hyper-parameter $\alpha$}
We investigate the sensitivity of the hyper-parameter $\alpha$ in the equation (\ref{stage1}), which balances the next-item prediction and cross-modality knowledge alignment loss. As shown in Figure \ref{fig:sensitivity of alpha}, we vary $\alpha$ from 0.2 to 5 and observe its effect on model performance.

Both Recall@20 and NDCG@20 improve significantly when $\alpha$ increases from 0.2 to 1.0, indicating that incorporating the alignment loss is beneficial for learning unified representations that integrate both collaborative and semantic information. Performance peaks when $\alpha$ is set to 1.0 or 2.0, suggesting that moderate emphasis on the alignment objective leads to optimal integration of modality-specific knowledge.

Importantly, the performance remains stable across a broad range of $\alpha$ values (from 0.5 to 5.0), demonstrating the robustness of the proposed method to this hyper-parameter. This insensitivity implies that our model is not overly reliant on fine-tuning $\alpha$, and can consistently benefit from the alignment signal as long as it is sufficiently incorporated. This robustness is particularly desirable in practical scenarios where extensive hyper-parameter search may be computationally prohibitive.

\begin{figure}[h]
  \centering
  \includegraphics[width=\linewidth]{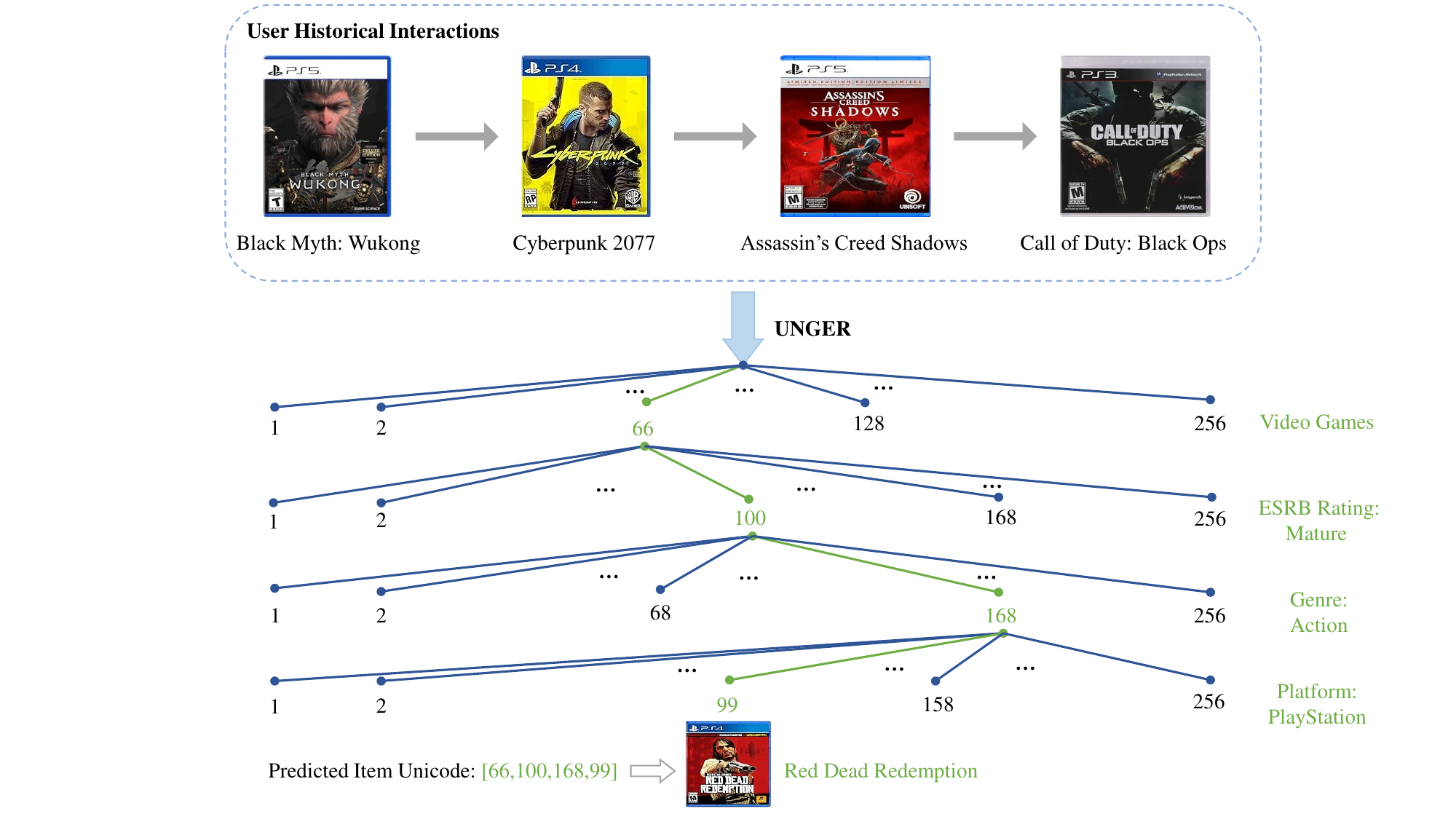}
  \caption{An example that illustrates the application of UNGER for the video game recommendation scenario.}
  \label{fig:case}
\end{figure}

\subsubsection{Sensitivity of hyper-parameter $\beta$}
We analyze the sensitivity of the hyper-parameter $\beta$ used in the equation (\ref{stage2}), which balances the generative recommendation objective and the intra-modality knowledge distillation task. As illustrated in Figure \ref{fig:sensitivity of beta}, we vary $\beta$ from 0.2 to 5.0 and evaluate the model performance.

Both Recall@20 and NDCG@20 initially improve as $\beta$ increases from 0.2 to 1.0, demonstrating the benefit of incorporating the distillation loss for enhancing the quality of the learned representations. This confirms that the global semantic signal distilled via the learnable token $c_{\text{dis}}$ provides complementary supervision that helps compensate for the information loss caused by quantization in Stage I.

The model achieves its best performance when $\beta = 1.0$, but remains stable across a relatively wide range (from 0.5 to 2.0), indicating that the method is robust to the choice of this hyper-parameter. Even when $\beta$ is increased to 5.0, performance does not deteriorate drastically, suggesting that the optimization process can effectively balance the two objectives. This robustness is particularly useful in practice, as it reduces the reliance on extensive hyper-parameter tuning during Stage II training.

\subsection{Case Study}
To demonstrate the interpretability and effectiveness of the proposed UNGER generative recommendation framework, we present a case study centered on personalized video game recommendation as shown in Figure \ref{fig:case}. This study highlights how UNGER performs autoregressive decoding over discrete \textit{unified codes}, which jointly encode semantic, behavioral, and contextual information to infer user intent and generate recommendations that align with multifaceted user preferences.

\subsubsection{User Historical Interaction Profiling}

We consider a representative user whose interaction history includes the following video games: \textit{Black Myth: Wukong}, \textit{Cyberpunk 2077}, \textit{Assassin’s Creed Shadows}, and \textit{Call of Duty: Black Ops}. These titles span multiple genres, narratives, and production styles. However, a pattern emerges when their key unified features are examined. The unified codes extracted from these items reflect consistent user affinities across several dimensions:

\begin{enumerate}
  \item \textbf{Gameplay Type and Genre:} Action-oriented experiences that involve combat, stealth, and open-world exploration.
  \item \textbf{Audience Targeting:} Mature content featuring violence, morally complex narratives, and sophisticated storytelling.
  \item \textbf{Platform Engagement:} A clear preference for PlayStation-compatible titles, suggesting a stable user-platform context.
\end{enumerate}

Unlike purely semantic models, UNGER does not rely on handcrafted or static attributes. Instead, each historical item is transformed into a sequence of discrete unified codes, learned through the training process to encapsulate relevant item-level and user-dependent information. These codes serve as the foundation for personalized sequence modeling.

\subsubsection{Autoregressive Unified Code Decoding}

Based on the unified codes derived from the user’s interaction history, UNGER proceeds with autoregressive decoding to infer the next most probable unified code sequence: $[66, 100, 168, 99]$. We first conduct a clustering analysis between the code distribution and item attributes, through which we find that certain codes consistently correspond to specific semantic concepts. Each code thus represents a learned latent concept that captures diverse patterns, such as item semantics, user intent, or platform preference. The decoding trajectory can be interpreted as follows:

\begin{enumerate}
  \item \textbf{Code 66 (Domain: Video Games).} Anchors the generation within the gaming domain, excluding items from other verticals such as music or movies.
  \item \textbf{Code 100 (Mature Content Affinity).} Indicates the model's inference of a mature content preference, derived from recurring patterns in prior game choices.
  \item \textbf{Code 168 (Action-Oriented Gameplay).} Captures the user’s interest in fast-paced, action-heavy interactions involving combat, tactics, or physical exploration.
  \item \textbf{Code 99 (Platform: PlayStation).} Implies a behavioral pattern involving consistent engagement with PlayStation-specific or compatible titles.
\end{enumerate}

While each code does not correspond directly to a human-defined label, their learned representations align with high-level concepts, allowing for interpretable projections. The unified code space thereby serves as a latent reasoning structure that guides generation.

\subsubsection{Recommendation Interpretation: \textit{Red Dead Redemption}}

The final recommendation produced by UNGER is \textit{Red Dead Redemption}. We map the generated unified code sequence to the profile of this title and find a high degree of alignment:

\begin{itemize}
  \item \textbf{Code 66 (Video Game Domain):} \textit{Red Dead Redemption} is firmly situated in the video game domain.
  \item \textbf{Code 100 (Mature Content):} The game’s ESRB rating is Mature, consistent with the user’s demonstrated comfort with adult themes.
  \item \textbf{Code 168 (Action-Adventure Gameplay):} The gameplay combines narrative, exploration, and action elements, mirroring previous interaction patterns.
  \item \textbf{Code 99 (Platform):} Available on PlayStation, the game respects the user’s device and ecosystem usage pattern.
\end{itemize}

Despite the fact that unified codes are not manually engineered semantic tags, their learned representations align well with interpretable item features, thereby enabling natural explanations post-hoc.

\subsubsection{Discussion and Insights}

This case study highlights UNGER’s capability to bridge implicit user behavior and explicit recommendation reasoning via a discrete code generation process. Compared to traditional recommendation systems that rely on uninterpretable latent embeddings or static feature matching, UNGER leverages unified code modeling to incrementally build a high-fidelity representation of user intent. The sequential decoding procedure refines the search space across multiple dimensions, which include domain, content style, gameplay genre, and platform, thus mimicking a step-by-step decision-making process similar to how users articulate preferences.

Moreover, the discrete and interpretable nature of the unified code space provides a transparent layer between user history and item selection. The recommendation of \textit{Red Dead Redemption} exemplifies how UNGER can reconstruct complex, multi-attribute user intent and map it to appropriate content, validating the framework's design objectives: transparency, personalization, and controllable generation.

\section{CONCLUSION AND FUTURE WORK} In this paper, we introduce a novel framework, UNGER, aimed at leveraging a unified code to encode knowledge from both semantic and collaborative modalities in generative recommendation systems. UNGER is composed of three key components: (1) A unified code generation framework that integrates semantic and collaborative knowledge, addressing the semantic domination issue by effectively capturing their complementary characteristics. (2) A joint optimization of cross-modality knowledge alignment and next item prediction tasks and a learnable modality adaption layer with AdaLN, which integrate collaborative and semantic knowledge into integrated embeddings, enabling the unified code to represent both modalities information cohesively. (3) An intra-modality knowledge distillation task with a specially designed token, which compensates for the information loss caused by the quantization process and further improves auto-regressive generation quality. Extensive experiments compared with state-of-the-art methods, combined with detailed analyses, validate the effectiveness and robustness of UNGER. 

In future work, we plan to extend UNGER along two complementary directions. The first direction involves enriching the semantic modality by incorporating visual embeddings extracted from item images or video frames using pretrained deep neural networks \cite{he2016deep, radford2021learning} and Vision Transformers \cite{dosovitskiy2020image}, as well as audio and speech representations from domains such as music and podcasts. This enhancement will enable the unified code to capture stylistic, aesthetic, and affective aspects of user preferences that are not fully conveyed by textual descriptions alone. The second direction focuses on improving the tokenizer by adopting advanced vector quantization techniques from other fields, including LFQ \cite{yu2023language}, IBQ \cite{shi2024taming}, and ActionPiece \cite{hou2025actionpiece}, which provide better convergence properties, higher codebook utilization, and more expressive latent representations. These improvements are expected to further enhance the generation quality and robustness of the model.

It is also worth noting that the current evaluation is restricted to relatively sparse subsets of the Amazon dataset. While this choice is consistent with prior work, it does imply that the behavior of UNGER on denser recommendation scenarios has not yet been systematically explored. We leave such evaluations as an important avenue for future investigation, which will help establish the broader applicability and generalizability of the proposed framework.

\begin{acks}
This work is supported by the National Key Research and Development Program of China under grant 2024YFC3307900; the National Natural Science Foundation of China under grants 62302184, 62376103, 62436003 and 62206102; Major Science and Technology Project of Hubei Province under grant 2024BAA008; Hubei Science and Technology Talent Service Project under grant 2024DJC078; Ant Group through CCF-Ant Research Fund. The computation is completed in the HPC Platform of Huazhong University of Science and Technology.  
\end{acks}


\bibliographystyle{ACM-Reference-Format}
\bibliography{reference}


\begin{thebibliography}{76}


\ifx \showCODEN    \undefined \def \showCODEN     #1{\unskip}     \fi
\ifx \showISBNx    \undefined \def \showISBNx     #1{\unskip}     \fi
\ifx \showISBNxiii \undefined \def \showISBNxiii  #1{\unskip}     \fi
\ifx \showISSN     \undefined \def \showISSN      #1{\unskip}     \fi
\ifx \showLCCN     \undefined \def \showLCCN      #1{\unskip}     \fi
\ifx \shownote     \undefined \def \shownote      #1{#1}          \fi
\ifx \showarticletitle \undefined \def \showarticletitle #1{#1}   \fi
\ifx \showURL      \undefined \def \showURL       {\relax}        \fi
\providecommand\bibfield[2]{#2}
\providecommand\bibinfo[2]{#2}
\providecommand\natexlab[1]{#1}
\providecommand\showeprint[2][]{arXiv:#2}

\bibitem[Andoni et~al\mbox{.}(2018)]%
        {andoni2018approximate}
\bibfield{author}{\bibinfo{person}{Alexandr Andoni}, \bibinfo{person}{Piotr Indyk}, {and} \bibinfo{person}{Ilya Razenshteyn}.} \bibinfo{year}{2018}\natexlab{}.
\newblock \showarticletitle{Approximate nearest neighbor search in high dimensions}. In \bibinfo{booktitle}{\emph{Proceedings of the International Congress of Mathematicians: Rio de Janeiro 2018}}. World Scientific, \bibinfo{pages}{3287--3318}.
\newblock


\bibitem[Arya et~al\mbox{.}(1998)]%
        {arya1998optimal}
\bibfield{author}{\bibinfo{person}{Sunil Arya}, \bibinfo{person}{David~M Mount}, \bibinfo{person}{Nathan~S Netanyahu}, \bibinfo{person}{Ruth Silverman}, {and} \bibinfo{person}{Angela~Y Wu}.} \bibinfo{year}{1998}\natexlab{}.
\newblock \showarticletitle{An optimal algorithm for approximate nearest neighbor searching fixed dimensions}.
\newblock \bibinfo{journal}{\emph{Journal of the ACM (JACM)}} \bibinfo{volume}{45}, \bibinfo{number}{6} (\bibinfo{year}{1998}), \bibinfo{pages}{891--923}.
\newblock


\bibitem[Chen et~al\mbox{.}(2020)]%
        {chen2020simple}
\bibfield{author}{\bibinfo{person}{Ting Chen}, \bibinfo{person}{Simon Kornblith}, \bibinfo{person}{Mohammad Norouzi}, {and} \bibinfo{person}{Geoffrey Hinton}.} \bibinfo{year}{2020}\natexlab{}.
\newblock \showarticletitle{A simple framework for contrastive learning of visual representations}. In \bibinfo{booktitle}{\emph{International conference on machine learning}}. PMLR, \bibinfo{pages}{1597--1607}.
\newblock


\bibitem[Cheng et~al\mbox{.}(2021)]%
        {cheng2021learning}
\bibfield{author}{\bibinfo{person}{Mingyue Cheng}, \bibinfo{person}{Fajie Yuan}, \bibinfo{person}{Qi Liu}, \bibinfo{person}{Xin Xin}, {and} \bibinfo{person}{Enhong Chen}.} \bibinfo{year}{2021}\natexlab{}.
\newblock \showarticletitle{Learning transferable user representations with sequential behaviors via contrastive pre-training}. In \bibinfo{booktitle}{\emph{2021 IEEE International Conference on Data Mining (ICDM)}}. IEEE, \bibinfo{pages}{51--60}.
\newblock


\bibitem[Covington et~al\mbox{.}(2016)]%
        {covington2016deep}
\bibfield{author}{\bibinfo{person}{Paul Covington}, \bibinfo{person}{Jay Adams}, {and} \bibinfo{person}{Emre Sargin}.} \bibinfo{year}{2016}\natexlab{}.
\newblock \showarticletitle{Deep neural networks for youtube recommendations}. In \bibinfo{booktitle}{\emph{Proceedings of the 10th ACM conference on recommender systems}}. \bibinfo{pages}{191--198}.
\newblock


\bibitem[Devlin(2018)]%
        {devlin2018bert}
\bibfield{author}{\bibinfo{person}{Jacob Devlin}.} \bibinfo{year}{2018}\natexlab{}.
\newblock \showarticletitle{Bert: Pre-training of deep bidirectional transformers for language understanding}.
\newblock \bibinfo{journal}{\emph{arXiv preprint arXiv:1810.04805}} (\bibinfo{year}{2018}).
\newblock


\bibitem[Devlin et~al\mbox{.}(2019)]%
        {devlin2019bert}
\bibfield{author}{\bibinfo{person}{Jacob Devlin}, \bibinfo{person}{Ming-Wei Chang}, \bibinfo{person}{Kenton Lee}, {and} \bibinfo{person}{Kristina Toutanova}.} \bibinfo{year}{2019}\natexlab{}.
\newblock \showarticletitle{Bert: Pre-training of deep bidirectional transformers for language understanding}. In \bibinfo{booktitle}{\emph{Proceedings of the 2019 conference of the North American chapter of the association for computational linguistics: human language technologies, volume 1 (long and short papers)}}. \bibinfo{pages}{4171--4186}.
\newblock


\bibitem[Diao et~al\mbox{.}(2014)]%
        {diao2014jointly}
\bibfield{author}{\bibinfo{person}{Qiming Diao}, \bibinfo{person}{Minghui Qiu}, \bibinfo{person}{Chao-Yuan Wu}, \bibinfo{person}{Alexander~J Smola}, \bibinfo{person}{Jing Jiang}, {and} \bibinfo{person}{Chong Wang}.} \bibinfo{year}{2014}\natexlab{}.
\newblock \showarticletitle{Jointly modeling aspects, ratings and sentiments for movie recommendation (JMARS)}. In \bibinfo{booktitle}{\emph{Proceedings of the 20th ACM SIGKDD international conference on Knowledge discovery and data mining}}. \bibinfo{pages}{193--202}.
\newblock


\bibitem[Ding et~al\mbox{.}(2021)]%
        {ding2021zero}
\bibfield{author}{\bibinfo{person}{Hao Ding}, \bibinfo{person}{Yifei Ma}, \bibinfo{person}{Anoop Deoras}, \bibinfo{person}{Yuyang Wang}, {and} \bibinfo{person}{Hao Wang}.} \bibinfo{year}{2021}\natexlab{}.
\newblock \showarticletitle{Zero-shot recommender systems}.
\newblock \bibinfo{journal}{\emph{arXiv preprint arXiv:2105.08318}} (\bibinfo{year}{2021}).
\newblock


\bibitem[Dosovitskiy et~al\mbox{.}(2020)]%
        {dosovitskiy2020image}
\bibfield{author}{\bibinfo{person}{Alexey Dosovitskiy}, \bibinfo{person}{Lucas Beyer}, \bibinfo{person}{Alexander Kolesnikov}, \bibinfo{person}{Dirk Weissenborn}, \bibinfo{person}{Xiaohua Zhai}, \bibinfo{person}{Thomas Unterthiner}, \bibinfo{person}{Mostafa Dehghani}, \bibinfo{person}{Matthias Minderer}, \bibinfo{person}{Georg Heigold}, \bibinfo{person}{Sylvain Gelly}, {et~al\mbox{.}}} \bibinfo{year}{2020}\natexlab{}.
\newblock \showarticletitle{An image is worth 16x16 words: Transformers for image recognition at scale}.
\newblock \bibinfo{journal}{\emph{arXiv preprint arXiv:2010.11929}} (\bibinfo{year}{2020}).
\newblock


\bibitem[Feng et~al\mbox{.}(2022)]%
        {feng2022recommender}
\bibfield{author}{\bibinfo{person}{Chao Feng}, \bibinfo{person}{Wuchao Li}, \bibinfo{person}{Defu Lian}, \bibinfo{person}{Zheng Liu}, {and} \bibinfo{person}{Enhong Chen}.} \bibinfo{year}{2022}\natexlab{}.
\newblock \showarticletitle{Recommender forest for efficient retrieval}.
\newblock \bibinfo{journal}{\emph{Advances in Neural Information Processing Systems}}  \bibinfo{volume}{35} (\bibinfo{year}{2022}), \bibinfo{pages}{38912--38924}.
\newblock


\bibitem[Geng et~al\mbox{.}(2022)]%
        {geng2022recommendation}
\bibfield{author}{\bibinfo{person}{Shijie Geng}, \bibinfo{person}{Shuchang Liu}, \bibinfo{person}{Zuohui Fu}, \bibinfo{person}{Yingqiang Ge}, {and} \bibinfo{person}{Yongfeng Zhang}.} \bibinfo{year}{2022}\natexlab{}.
\newblock \showarticletitle{Recommendation as language processing (rlp): A unified pretrain, personalized prompt \& predict paradigm (p5)}. In \bibinfo{booktitle}{\emph{Proceedings of the 16th ACM conference on recommender systems}}. \bibinfo{pages}{299--315}.
\newblock


\bibitem[Guo et~al\mbox{.}(2025)]%
        {guo2025deepseek}
\bibfield{author}{\bibinfo{person}{Daya Guo}, \bibinfo{person}{Dejian Yang}, \bibinfo{person}{Haowei Zhang}, \bibinfo{person}{Junxiao Song}, \bibinfo{person}{Ruoyu Zhang}, \bibinfo{person}{Runxin Xu}, \bibinfo{person}{Qihao Zhu}, \bibinfo{person}{Shirong Ma}, \bibinfo{person}{Peiyi Wang}, \bibinfo{person}{Xiao Bi}, {et~al\mbox{.}}} \bibinfo{year}{2025}\natexlab{}.
\newblock \showarticletitle{Deepseek-r1: Incentivizing reasoning capability in llms via reinforcement learning}.
\newblock \bibinfo{journal}{\emph{arXiv preprint arXiv:2501.12948}} (\bibinfo{year}{2025}).
\newblock


\bibitem[Guo et~al\mbox{.}(2020)]%
        {guo2020accelerating}
\bibfield{author}{\bibinfo{person}{Ruiqi Guo}, \bibinfo{person}{Philip Sun}, \bibinfo{person}{Erik Lindgren}, \bibinfo{person}{Quan Geng}, \bibinfo{person}{David Simcha}, \bibinfo{person}{Felix Chern}, {and} \bibinfo{person}{Sanjiv Kumar}.} \bibinfo{year}{2020}\natexlab{}.
\newblock \showarticletitle{Accelerating large-scale inference with anisotropic vector quantization}. In \bibinfo{booktitle}{\emph{International Conference on Machine Learning}}. PMLR, \bibinfo{pages}{3887--3896}.
\newblock


\bibitem[He et~al\mbox{.}(2016)]%
        {he2016deep}
\bibfield{author}{\bibinfo{person}{Kaiming He}, \bibinfo{person}{Xiangyu Zhang}, \bibinfo{person}{Shaoqing Ren}, {and} \bibinfo{person}{Jian Sun}.} \bibinfo{year}{2016}\natexlab{}.
\newblock \showarticletitle{Deep residual learning for image recognition}. In \bibinfo{booktitle}{\emph{Proceedings of the IEEE conference on computer vision and pattern recognition}}. \bibinfo{pages}{770--778}.
\newblock


\bibitem[He and McAuley(2016)]%
        {he2016fusing}
\bibfield{author}{\bibinfo{person}{Ruining He} {and} \bibinfo{person}{Julian McAuley}.} \bibinfo{year}{2016}\natexlab{}.
\newblock \showarticletitle{Fusing similarity models with markov chains for sparse sequential recommendation}. In \bibinfo{booktitle}{\emph{2016 IEEE 16th international conference on data mining (ICDM)}}. IEEE, \bibinfo{pages}{191--200}.
\newblock


\bibitem[Hidasi(2015)]%
        {hidasi2015session}
\bibfield{author}{\bibinfo{person}{B Hidasi}.} \bibinfo{year}{2015}\natexlab{}.
\newblock \showarticletitle{Session-based Recommendations with Recurrent Neural Networks}.
\newblock \bibinfo{journal}{\emph{arXiv preprint arXiv:1511.06939}} (\bibinfo{year}{2015}).
\newblock


\bibitem[Hou et~al\mbox{.}(2022)]%
        {hou2022towards}
\bibfield{author}{\bibinfo{person}{Yupeng Hou}, \bibinfo{person}{Shanlei Mu}, \bibinfo{person}{Wayne~Xin Zhao}, \bibinfo{person}{Yaliang Li}, \bibinfo{person}{Bolin Ding}, {and} \bibinfo{person}{Ji-Rong Wen}.} \bibinfo{year}{2022}\natexlab{}.
\newblock \showarticletitle{Towards universal sequence representation learning for recommender systems}. In \bibinfo{booktitle}{\emph{Proceedings of the 28th ACM SIGKDD Conference on Knowledge Discovery and Data Mining}}. \bibinfo{pages}{585--593}.
\newblock


\bibitem[Hou et~al\mbox{.}(2025)]%
        {hou2025actionpiece}
\bibfield{author}{\bibinfo{person}{Yupeng Hou}, \bibinfo{person}{Jianmo Ni}, \bibinfo{person}{Zhankui He}, \bibinfo{person}{Noveen Sachdeva}, \bibinfo{person}{Wang-Cheng Kang}, \bibinfo{person}{Ed~H Chi}, \bibinfo{person}{Julian McAuley}, {and} \bibinfo{person}{Derek~Zhiyuan Cheng}.} \bibinfo{year}{2025}\natexlab{}.
\newblock \showarticletitle{ActionPiece: Contextually Tokenizing Action Sequences for Generative Recommendation}.
\newblock \bibinfo{journal}{\emph{arXiv preprint arXiv:2502.13581}} (\bibinfo{year}{2025}).
\newblock


\bibitem[Hurst et~al\mbox{.}(2024)]%
        {hurst2024gpt}
\bibfield{author}{\bibinfo{person}{Aaron Hurst}, \bibinfo{person}{Adam Lerer}, \bibinfo{person}{Adam~P Goucher}, \bibinfo{person}{Adam Perelman}, \bibinfo{person}{Aditya Ramesh}, \bibinfo{person}{Aidan Clark}, \bibinfo{person}{AJ Ostrow}, \bibinfo{person}{Akila Welihinda}, \bibinfo{person}{Alan Hayes}, \bibinfo{person}{Alec Radford}, {et~al\mbox{.}}} \bibinfo{year}{2024}\natexlab{}.
\newblock \showarticletitle{Gpt-4o system card}.
\newblock \bibinfo{journal}{\emph{arXiv preprint arXiv:2410.21276}} (\bibinfo{year}{2024}).
\newblock


\bibitem[Jannach and Ludewig(2017)]%
        {jannach2017recurrent}
\bibfield{author}{\bibinfo{person}{Dietmar Jannach} {and} \bibinfo{person}{Malte Ludewig}.} \bibinfo{year}{2017}\natexlab{}.
\newblock \showarticletitle{When recurrent neural networks meet the neighborhood for session-based recommendation}. In \bibinfo{booktitle}{\emph{Proceedings of the eleventh ACM conference on recommender systems}}. \bibinfo{pages}{306--310}.
\newblock


\bibitem[Ji et~al\mbox{.}(2024)]%
        {ji2024genrec}
\bibfield{author}{\bibinfo{person}{Jianchao Ji}, \bibinfo{person}{Zelong Li}, \bibinfo{person}{Shuyuan Xu}, \bibinfo{person}{Wenyue Hua}, \bibinfo{person}{Yingqiang Ge}, \bibinfo{person}{Juntao Tan}, {and} \bibinfo{person}{Yongfeng Zhang}.} \bibinfo{year}{2024}\natexlab{}.
\newblock \showarticletitle{Genrec: Large language model for generative recommendation}. In \bibinfo{booktitle}{\emph{European Conference on Information Retrieval}}. Springer, \bibinfo{pages}{494--502}.
\newblock


\bibitem[Johnson et~al\mbox{.}(2019)]%
        {johnson2019billion}
\bibfield{author}{\bibinfo{person}{Jeff Johnson}, \bibinfo{person}{Matthijs Douze}, {and} \bibinfo{person}{Herv{\'e} J{\'e}gou}.} \bibinfo{year}{2019}\natexlab{}.
\newblock \showarticletitle{Billion-scale similarity search with GPUs}.
\newblock \bibinfo{journal}{\emph{IEEE Transactions on Big Data}} \bibinfo{volume}{7}, \bibinfo{number}{3} (\bibinfo{year}{2019}), \bibinfo{pages}{535--547}.
\newblock


\bibitem[Kang and McAuley(2018)]%
        {kang2018self}
\bibfield{author}{\bibinfo{person}{Wang-Cheng Kang} {and} \bibinfo{person}{Julian McAuley}.} \bibinfo{year}{2018}\natexlab{}.
\newblock \showarticletitle{Self-attentive sequential recommendation}. In \bibinfo{booktitle}{\emph{2018 IEEE international conference on data mining (ICDM)}}. IEEE, \bibinfo{pages}{197--206}.
\newblock


\bibitem[Kim et~al\mbox{.}(2007)]%
        {kim2007music}
\bibfield{author}{\bibinfo{person}{Dongmoon Kim}, \bibinfo{person}{Kun-su Kim}, \bibinfo{person}{Kyo-Hyun Park}, \bibinfo{person}{Jee-Hyong Lee}, {and} \bibinfo{person}{Keon~Myung Lee}.} \bibinfo{year}{2007}\natexlab{}.
\newblock \showarticletitle{A music recommendation system with a dynamic k-means clustering algorithm}. In \bibinfo{booktitle}{\emph{Sixth international conference on machine learning and applications (ICMLA 2007)}}. IEEE, \bibinfo{pages}{399--403}.
\newblock


\bibitem[Kim et~al\mbox{.}(2024)]%
        {kim2024sc}
\bibfield{author}{\bibinfo{person}{Tongyoung Kim}, \bibinfo{person}{Soojin Yoon}, \bibinfo{person}{Seongku Kang}, \bibinfo{person}{Jinyoung Yeo}, {and} \bibinfo{person}{Dongha Lee}.} \bibinfo{year}{2024}\natexlab{}.
\newblock \showarticletitle{SC-Rec: Enhancing Generative Retrieval with Self-Consistent Reranking for Sequential Recommendation}.
\newblock \bibinfo{journal}{\emph{arXiv preprint arXiv:2408.08686}} (\bibinfo{year}{2024}).
\newblock


\bibitem[Lam et~al\mbox{.}(2008)]%
        {lam2008addressing}
\bibfield{author}{\bibinfo{person}{Xuan~Nhat Lam}, \bibinfo{person}{Thuc Vu}, \bibinfo{person}{Trong~Duc Le}, {and} \bibinfo{person}{Anh~Duc Duong}.} \bibinfo{year}{2008}\natexlab{}.
\newblock \showarticletitle{Addressing cold-start problem in recommendation systems}. In \bibinfo{booktitle}{\emph{Proceedings of the 2nd international conference on Ubiquitous information management and communication}}. \bibinfo{pages}{208--211}.
\newblock


\bibitem[Lee et~al\mbox{.}(2022)]%
        {lee2022autoregressive}
\bibfield{author}{\bibinfo{person}{Doyup Lee}, \bibinfo{person}{Chiheon Kim}, \bibinfo{person}{Saehoon Kim}, \bibinfo{person}{Minsu Cho}, {and} \bibinfo{person}{Wook-Shin Han}.} \bibinfo{year}{2022}\natexlab{}.
\newblock \showarticletitle{Autoregressive image generation using residual quantization}. In \bibinfo{booktitle}{\emph{Proceedings of the IEEE/CVF Conference on Computer Vision and Pattern Recognition}}. \bibinfo{pages}{11523--11532}.
\newblock


\bibitem[Lee and Toutanova(2018)]%
        {lee2018pre}
\bibfield{author}{\bibinfo{person}{JDMCK Lee} {and} \bibinfo{person}{K Toutanova}.} \bibinfo{year}{2018}\natexlab{}.
\newblock \showarticletitle{Pre-training of deep bidirectional transformers for language understanding}.
\newblock \bibinfo{journal}{\emph{arXiv preprint arXiv:1810.04805}} \bibinfo{volume}{3}, \bibinfo{number}{8} (\bibinfo{year}{2018}).
\newblock


\bibitem[Li et~al\mbox{.}(2022)]%
        {li2022blip}
\bibfield{author}{\bibinfo{person}{Junnan Li}, \bibinfo{person}{Dongxu Li}, \bibinfo{person}{Caiming Xiong}, {and} \bibinfo{person}{Steven Hoi}.} \bibinfo{year}{2022}\natexlab{}.
\newblock \showarticletitle{Blip: Bootstrapping language-image pre-training for unified vision-language understanding and generation}. In \bibinfo{booktitle}{\emph{International conference on machine learning}}. PMLR, \bibinfo{pages}{12888--12900}.
\newblock


\bibitem[Li et~al\mbox{.}(2024)]%
        {li2024embedding}
\bibfield{author}{\bibinfo{person}{Shiwei Li}, \bibinfo{person}{Huifeng Guo}, \bibinfo{person}{Xing Tang}, \bibinfo{person}{Ruiming Tang}, \bibinfo{person}{Lu Hou}, \bibinfo{person}{Ruixuan Li}, {and} \bibinfo{person}{Rui Zhang}.} \bibinfo{year}{2024}\natexlab{}.
\newblock \showarticletitle{Embedding compression in recommender systems: A survey}.
\newblock \bibinfo{journal}{\emph{Comput. Surveys}} \bibinfo{volume}{56}, \bibinfo{number}{5} (\bibinfo{year}{2024}), \bibinfo{pages}{1--21}.
\newblock


\bibitem[Li et~al\mbox{.}(2025)]%
        {li2025personalized}
\bibfield{author}{\bibinfo{person}{Yichen Li}, \bibinfo{person}{Yijing Shan}, \bibinfo{person}{Yi Liu}, \bibinfo{person}{Haozhao Wang}, \bibinfo{person}{Wei Wang}, \bibinfo{person}{Yi Wang}, {and} \bibinfo{person}{Ruixuan Li}.} \bibinfo{year}{2025}\natexlab{}.
\newblock \showarticletitle{Personalized Federated Recommendation for Cold-Start Users via Adaptive Knowledge Fusion}. In \bibinfo{booktitle}{\emph{Proceedings of the ACM on Web Conference 2025}}. \bibinfo{pages}{2700--2709}.
\newblock


\bibitem[Lika et~al\mbox{.}(2014)]%
        {lika2014facing}
\bibfield{author}{\bibinfo{person}{Blerina Lika}, \bibinfo{person}{Kostas Kolomvatsos}, {and} \bibinfo{person}{Stathes Hadjiefthymiades}.} \bibinfo{year}{2014}\natexlab{}.
\newblock \showarticletitle{Facing the cold start problem in recommender systems}.
\newblock \bibinfo{journal}{\emph{Expert systems with applications}} \bibinfo{volume}{41}, \bibinfo{number}{4} (\bibinfo{year}{2014}), \bibinfo{pages}{2065--2073}.
\newblock


\bibitem[Lin et~al\mbox{.}(2024)]%
        {lin2024bridging}
\bibfield{author}{\bibinfo{person}{Xinyu Lin}, \bibinfo{person}{Wenjie Wang}, \bibinfo{person}{Yongqi Li}, \bibinfo{person}{Fuli Feng}, \bibinfo{person}{See-Kiong Ng}, {and} \bibinfo{person}{Tat-Seng Chua}.} \bibinfo{year}{2024}\natexlab{}.
\newblock \showarticletitle{Bridging items and language: A transition paradigm for large language model-based recommendation}. In \bibinfo{booktitle}{\emph{Proceedings of the 30th ACM SIGKDD Conference on Knowledge Discovery and Data Mining}}. \bibinfo{pages}{1816--1826}.
\newblock


\bibitem[Liu et~al\mbox{.}(2023)]%
        {liu2023visual}
\bibfield{author}{\bibinfo{person}{Haotian Liu}, \bibinfo{person}{Chunyuan Li}, \bibinfo{person}{Qingyang Wu}, {and} \bibinfo{person}{Yong~Jae Lee}.} \bibinfo{year}{2023}\natexlab{}.
\newblock \showarticletitle{Visual instruction tuning}.
\newblock \bibinfo{journal}{\emph{Advances in neural information processing systems}}  \bibinfo{volume}{36} (\bibinfo{year}{2023}), \bibinfo{pages}{34892--34916}.
\newblock


\bibitem[Liu et~al\mbox{.}(2024a)]%
        {liu2024mmgrec}
\bibfield{author}{\bibinfo{person}{Han Liu}, \bibinfo{person}{Yinwei Wei}, \bibinfo{person}{Xuemeng Song}, \bibinfo{person}{Weili Guan}, \bibinfo{person}{Yuan-Fang Li}, {and} \bibinfo{person}{Liqiang Nie}.} \bibinfo{year}{2024}\natexlab{a}.
\newblock \showarticletitle{MMGRec: Multimodal Generative Recommendation with Transformer Model}.
\newblock \bibinfo{journal}{\emph{arXiv preprint arXiv:2404.16555}} (\bibinfo{year}{2024}).
\newblock


\bibitem[Liu et~al\mbox{.}(2024c)]%
        {liu2024multimodal}
\bibfield{author}{\bibinfo{person}{Qijiong Liu}, \bibinfo{person}{Jieming Zhu}, \bibinfo{person}{Yanting Yang}, \bibinfo{person}{Quanyu Dai}, \bibinfo{person}{Zhaocheng Du}, \bibinfo{person}{Xiao-Ming Wu}, \bibinfo{person}{Zhou Zhao}, \bibinfo{person}{Rui Zhang}, {and} \bibinfo{person}{Zhenhua Dong}.} \bibinfo{year}{2024}\natexlab{c}.
\newblock \showarticletitle{Multimodal pretraining, adaptation, and generation for recommendation: A survey}. In \bibinfo{booktitle}{\emph{Proceedings of the 30th ACM SIGKDD Conference on Knowledge Discovery and Data Mining}}. \bibinfo{pages}{6566--6576}.
\newblock


\bibitem[Liu et~al\mbox{.}(2024b)]%
        {liu2024collaborative}
\bibfield{author}{\bibinfo{person}{Zhongzhou Liu}, \bibinfo{person}{Hao Zhang}, \bibinfo{person}{Kuicai Dong}, {and} \bibinfo{person}{Yuan Fang}.} \bibinfo{year}{2024}\natexlab{b}.
\newblock \showarticletitle{Collaborative Cross-modal Fusion with Large Language Model for Recommendation}. In \bibinfo{booktitle}{\emph{Proceedings of the 33rd ACM International Conference on Information and Knowledge Management}}. \bibinfo{pages}{1565--1574}.
\newblock


\bibitem[Ma et~al\mbox{.}(2019)]%
        {ma2019hierarchical}
\bibfield{author}{\bibinfo{person}{Chen Ma}, \bibinfo{person}{Peng Kang}, {and} \bibinfo{person}{Xue Liu}.} \bibinfo{year}{2019}\natexlab{}.
\newblock \showarticletitle{Hierarchical gating networks for sequential recommendation}. In \bibinfo{booktitle}{\emph{Proceedings of the 25th ACM SIGKDD international conference on knowledge discovery \& data mining}}. \bibinfo{pages}{825--833}.
\newblock


\bibitem[McAuley et~al\mbox{.}(2015)]%
        {mcauley2015image}
\bibfield{author}{\bibinfo{person}{Julian McAuley}, \bibinfo{person}{Christopher Targett}, \bibinfo{person}{Qinfeng Shi}, {and} \bibinfo{person}{Anton Van Den~Hengel}.} \bibinfo{year}{2015}\natexlab{}.
\newblock \showarticletitle{Image-based recommendations on styles and substitutes}. In \bibinfo{booktitle}{\emph{Proceedings of the 38th international ACM SIGIR conference on research and development in information retrieval}}. \bibinfo{pages}{43--52}.
\newblock


\bibitem[Ni et~al\mbox{.}(2021)]%
        {ni2021sentence}
\bibfield{author}{\bibinfo{person}{Jianmo Ni}, \bibinfo{person}{Gustavo~Hernandez Abrego}, \bibinfo{person}{Noah Constant}, \bibinfo{person}{Ji Ma}, \bibinfo{person}{Keith~B Hall}, \bibinfo{person}{Daniel Cer}, {and} \bibinfo{person}{Yinfei Yang}.} \bibinfo{year}{2021}\natexlab{}.
\newblock \showarticletitle{Sentence-t5: Scalable sentence encoders from pre-trained text-to-text models}.
\newblock \bibinfo{journal}{\emph{arXiv preprint arXiv:2108.08877}} (\bibinfo{year}{2021}).
\newblock


\bibitem[Pan et~al\mbox{.}(2019)]%
        {pan2019warm}
\bibfield{author}{\bibinfo{person}{Feiyang Pan}, \bibinfo{person}{Shuokai Li}, \bibinfo{person}{Xiang Ao}, \bibinfo{person}{Pingzhong Tang}, {and} \bibinfo{person}{Qing He}.} \bibinfo{year}{2019}\natexlab{}.
\newblock \showarticletitle{Warm up cold-start advertisements: Improving ctr predictions via learning to learn id embeddings}. In \bibinfo{booktitle}{\emph{Proceedings of the 42nd International ACM SIGIR Conference on Research and Development in Information Retrieval}}. \bibinfo{pages}{695--704}.
\newblock


\bibitem[Peebles and Xie(2023)]%
        {peebles2023scalable}
\bibfield{author}{\bibinfo{person}{William Peebles} {and} \bibinfo{person}{Saining Xie}.} \bibinfo{year}{2023}\natexlab{}.
\newblock \showarticletitle{Scalable diffusion models with transformers}. In \bibinfo{booktitle}{\emph{Proceedings of the IEEE/CVF International Conference on Computer Vision}}. \bibinfo{pages}{4195--4205}.
\newblock


\bibitem[Peng et~al\mbox{.}(2022)]%
        {peng2022balanced}
\bibfield{author}{\bibinfo{person}{Xiaokang Peng}, \bibinfo{person}{Yake Wei}, \bibinfo{person}{Andong Deng}, \bibinfo{person}{Dong Wang}, {and} \bibinfo{person}{Di Hu}.} \bibinfo{year}{2022}\natexlab{}.
\newblock \showarticletitle{Balanced multimodal learning via on-the-fly gradient modulation}. In \bibinfo{booktitle}{\emph{Proceedings of the IEEE/CVF conference on computer vision and pattern recognition}}. \bibinfo{pages}{8238--8247}.
\newblock


\bibitem[Qi et~al\mbox{.}(2017)]%
        {qi2017effective}
\bibfield{author}{\bibinfo{person}{Jianpeng Qi}, \bibinfo{person}{Yanwei Yu}, \bibinfo{person}{Lihong Wang}, \bibinfo{person}{Jinglei Liu}, {and} \bibinfo{person}{Yingjie Wang}.} \bibinfo{year}{2017}\natexlab{}.
\newblock \showarticletitle{An effective and efficient hierarchical K-means clustering algorithm}.
\newblock \bibinfo{journal}{\emph{International Journal of Distributed Sensor Networks}} \bibinfo{volume}{13}, \bibinfo{number}{8} (\bibinfo{year}{2017}), \bibinfo{pages}{1550147717728627}.
\newblock


\bibitem[Radford et~al\mbox{.}(2021)]%
        {radford2021learning}
\bibfield{author}{\bibinfo{person}{Alec Radford}, \bibinfo{person}{Jong~Wook Kim}, \bibinfo{person}{Chris Hallacy}, \bibinfo{person}{Aditya Ramesh}, \bibinfo{person}{Gabriel Goh}, \bibinfo{person}{Sandhini Agarwal}, \bibinfo{person}{Girish Sastry}, \bibinfo{person}{Amanda Askell}, \bibinfo{person}{Pamela Mishkin}, \bibinfo{person}{Jack Clark}, {et~al\mbox{.}}} \bibinfo{year}{2021}\natexlab{}.
\newblock \showarticletitle{Learning transferable visual models from natural language supervision}. In \bibinfo{booktitle}{\emph{International conference on machine learning}}. PmLR, \bibinfo{pages}{8748--8763}.
\newblock


\bibitem[Rajput et~al\mbox{.}(2023)]%
        {rajput2023recommender}
\bibfield{author}{\bibinfo{person}{Shashank Rajput}, \bibinfo{person}{Nikhil Mehta}, \bibinfo{person}{Anima Singh}, \bibinfo{person}{Raghunandan Hulikal~Keshavan}, \bibinfo{person}{Trung Vu}, \bibinfo{person}{Lukasz Heldt}, \bibinfo{person}{Lichan Hong}, \bibinfo{person}{Yi Tay}, \bibinfo{person}{Vinh Tran}, \bibinfo{person}{Jonah Samost}, {et~al\mbox{.}}} \bibinfo{year}{2023}\natexlab{}.
\newblock \showarticletitle{Recommender systems with generative retrieval}.
\newblock \bibinfo{journal}{\emph{Advances in Neural Information Processing Systems}}  \bibinfo{volume}{36} (\bibinfo{year}{2023}), \bibinfo{pages}{10299--10315}.
\newblock


\bibitem[Rendle et~al\mbox{.}(2010)]%
        {rendle2010factorizing}
\bibfield{author}{\bibinfo{person}{Steffen Rendle}, \bibinfo{person}{Christoph Freudenthaler}, {and} \bibinfo{person}{Lars Schmidt-Thieme}.} \bibinfo{year}{2010}\natexlab{}.
\newblock \showarticletitle{Factorizing personalized markov chains for next-basket recommendation}. In \bibinfo{booktitle}{\emph{Proceedings of the 19th international conference on World wide web}}. \bibinfo{pages}{811--820}.
\newblock


\bibitem[Shen et~al\mbox{.}(2024)]%
        {shen2024pmg}
\bibfield{author}{\bibinfo{person}{Xiaoteng Shen}, \bibinfo{person}{Rui Zhang}, \bibinfo{person}{Xiaoyan Zhao}, \bibinfo{person}{Jieming Zhu}, {and} \bibinfo{person}{Xi Xiao}.} \bibinfo{year}{2024}\natexlab{}.
\newblock \showarticletitle{Pmg: Personalized multimodal generation with large language models}. In \bibinfo{booktitle}{\emph{Proceedings of the ACM Web Conference 2024}}. \bibinfo{pages}{3833--3843}.
\newblock


\bibitem[Shi et~al\mbox{.}(2024)]%
        {shi2024taming}
\bibfield{author}{\bibinfo{person}{Fengyuan Shi}, \bibinfo{person}{Zhuoyan Luo}, \bibinfo{person}{Yixiao Ge}, \bibinfo{person}{Yujiu Yang}, \bibinfo{person}{Ying Shan}, {and} \bibinfo{person}{Limin Wang}.} \bibinfo{year}{2024}\natexlab{}.
\newblock \showarticletitle{Taming scalable visual tokenizer for autoregressive image generation}.
\newblock \bibinfo{journal}{\emph{arXiv preprint arXiv:2412.02692}} (\bibinfo{year}{2024}).
\newblock


\bibitem[Sun et~al\mbox{.}(2019)]%
        {sun2019bert4rec}
\bibfield{author}{\bibinfo{person}{Fei Sun}, \bibinfo{person}{Jun Liu}, \bibinfo{person}{Jian Wu}, \bibinfo{person}{Changhua Pei}, \bibinfo{person}{Xiao Lin}, \bibinfo{person}{Wenwu Ou}, {and} \bibinfo{person}{Peng Jiang}.} \bibinfo{year}{2019}\natexlab{}.
\newblock \showarticletitle{BERT4Rec: Sequential recommendation with bidirectional encoder representations from transformer}. In \bibinfo{booktitle}{\emph{Proceedings of the 28th ACM international conference on information and knowledge management}}. \bibinfo{pages}{1441--1450}.
\newblock


\bibitem[Tang and Wang(2018)]%
        {tang2018personalized}
\bibfield{author}{\bibinfo{person}{Jiaxi Tang} {and} \bibinfo{person}{Ke Wang}.} \bibinfo{year}{2018}\natexlab{}.
\newblock \showarticletitle{Personalized top-n sequential recommendation via convolutional sequence embedding}. In \bibinfo{booktitle}{\emph{Proceedings of the eleventh ACM international conference on web search and data mining}}. \bibinfo{pages}{565--573}.
\newblock


\bibitem[Touvron et~al\mbox{.}(2023)]%
        {touvron2023llama}
\bibfield{author}{\bibinfo{person}{Hugo Touvron}, \bibinfo{person}{Louis Martin}, \bibinfo{person}{Kevin Stone}, \bibinfo{person}{Peter Albert}, \bibinfo{person}{Amjad Almahairi}, \bibinfo{person}{Yasmine Babaei}, \bibinfo{person}{Nikolay Bashlykov}, \bibinfo{person}{Soumya Batra}, \bibinfo{person}{Prajjwal Bhargava}, \bibinfo{person}{Shruti Bhosale}, {et~al\mbox{.}}} \bibinfo{year}{2023}\natexlab{}.
\newblock \showarticletitle{Llama 2: Open foundation and fine-tuned chat models}.
\newblock \bibinfo{journal}{\emph{arXiv preprint arXiv:2307.09288}} (\bibinfo{year}{2023}).
\newblock


\bibitem[Wang et~al\mbox{.}({[n.\,d.]})]%
        {DBLP:conf/iclr/WangXLX0024}
\bibfield{author}{\bibinfo{person}{Haozhao Wang}, \bibinfo{person}{Haoran Xu}, \bibinfo{person}{Yichen Li}, \bibinfo{person}{Yuan Xu}, \bibinfo{person}{Ruixuan Li}, {and} \bibinfo{person}{Tianwei Zhang}.} \bibinfo{year}{[n.\,d.]}\natexlab{}.
\newblock \showarticletitle{FedCDA: Federated Learning with Cross-rounds Divergence-aware Aggregation}. In \bibinfo{booktitle}{\emph{The Twelfth International Conference on Learning Representations, {ICLR} 2024, Vienna, Austria, May 7-11, 2024}}.
\newblock


\bibitem[Wang et~al\mbox{.}(2018)]%
        {wang2018billion}
\bibfield{author}{\bibinfo{person}{Jizhe Wang}, \bibinfo{person}{Pipei Huang}, \bibinfo{person}{Huan Zhao}, \bibinfo{person}{Zhibo Zhang}, \bibinfo{person}{Binqiang Zhao}, {and} \bibinfo{person}{Dik~Lun Lee}.} \bibinfo{year}{2018}\natexlab{}.
\newblock \showarticletitle{Billion-scale commodity embedding for e-commerce recommendation in alibaba}. In \bibinfo{booktitle}{\emph{Proceedings of the 24th ACM SIGKDD international conference on knowledge discovery \& data mining}}. \bibinfo{pages}{839--848}.
\newblock


\bibitem[Wang et~al\mbox{.}(2023c)]%
        {wang2023missrec}
\bibfield{author}{\bibinfo{person}{Jinpeng Wang}, \bibinfo{person}{Ziyun Zeng}, \bibinfo{person}{Yunxiao Wang}, \bibinfo{person}{Yuting Wang}, \bibinfo{person}{Xingyu Lu}, \bibinfo{person}{Tianxiang Li}, \bibinfo{person}{Jun Yuan}, \bibinfo{person}{Rui Zhang}, \bibinfo{person}{Hai-Tao Zheng}, {and} \bibinfo{person}{Shu-Tao Xia}.} \bibinfo{year}{2023}\natexlab{c}.
\newblock \showarticletitle{Missrec: Pre-training and transferring multi-modal interest-aware sequence representation for recommendation}. In \bibinfo{booktitle}{\emph{Proceedings of the 31st ACM International Conference on Multimedia}}. \bibinfo{pages}{6548--6557}.
\newblock


\bibitem[Wang et~al\mbox{.}(2021)]%
        {wang2021cross}
\bibfield{author}{\bibinfo{person}{Jinpeng Wang}, \bibinfo{person}{Jieming Zhu}, {and} \bibinfo{person}{Xiuqiang He}.} \bibinfo{year}{2021}\natexlab{}.
\newblock \showarticletitle{Cross-batch negative sampling for training two-tower recommenders}. In \bibinfo{booktitle}{\emph{Proceedings of the 44th international ACM SIGIR conference on research and development in information retrieval}}. \bibinfo{pages}{1632--1636}.
\newblock


\bibitem[Wang et~al\mbox{.}(2024a)]%
        {wang2024learnable}
\bibfield{author}{\bibinfo{person}{Wenjie Wang}, \bibinfo{person}{Honghui Bao}, \bibinfo{person}{Xinyu Lin}, \bibinfo{person}{Jizhi Zhang}, \bibinfo{person}{Yongqi Li}, \bibinfo{person}{Fuli Feng}, \bibinfo{person}{See-Kiong Ng}, {and} \bibinfo{person}{Tat-Seng Chua}.} \bibinfo{year}{2024}\natexlab{a}.
\newblock \showarticletitle{Learnable item tokenization for generative recommendation}. In \bibinfo{booktitle}{\emph{Proceedings of the 33rd ACM International Conference on Information and Knowledge Management}}. \bibinfo{pages}{2400--2409}.
\newblock


\bibitem[Wang et~al\mbox{.}(2023a)]%
        {wang2023generative}
\bibfield{author}{\bibinfo{person}{Wenjie Wang}, \bibinfo{person}{Xinyu Lin}, \bibinfo{person}{Fuli Feng}, \bibinfo{person}{Xiangnan He}, {and} \bibinfo{person}{Tat-Seng Chua}.} \bibinfo{year}{2023}\natexlab{a}.
\newblock \showarticletitle{Generative recommendation: Towards next-generation recommender paradigm}.
\newblock \bibinfo{journal}{\emph{arXiv preprint arXiv:2304.03516}} (\bibinfo{year}{2023}).
\newblock


\bibitem[Wang et~al\mbox{.}(2023b)]%
        {wang2023diffusion}
\bibfield{author}{\bibinfo{person}{Wenjie Wang}, \bibinfo{person}{Yiyan Xu}, \bibinfo{person}{Fuli Feng}, \bibinfo{person}{Xinyu Lin}, \bibinfo{person}{Xiangnan He}, {and} \bibinfo{person}{Tat-Seng Chua}.} \bibinfo{year}{2023}\natexlab{b}.
\newblock \showarticletitle{Diffusion recommender model}. In \bibinfo{booktitle}{\emph{Proceedings of the 46th International ACM SIGIR Conference on Research and Development in Information Retrieval}}. \bibinfo{pages}{832--841}.
\newblock


\bibitem[Wang et~al\mbox{.}(2024b)]%
        {wang2024enhanced}
\bibfield{author}{\bibinfo{person}{Yidan Wang}, \bibinfo{person}{Zhaochun Ren}, \bibinfo{person}{Weiwei Sun}, \bibinfo{person}{Jiyuan Yang}, \bibinfo{person}{Zhixiang Liang}, \bibinfo{person}{Xin Chen}, \bibinfo{person}{Ruobing Xie}, \bibinfo{person}{Su Yan}, \bibinfo{person}{Xu Zhang}, \bibinfo{person}{Pengjie Ren}, {et~al\mbox{.}}} \bibinfo{year}{2024}\natexlab{b}.
\newblock \showarticletitle{Enhanced generative recommendation via content and collaboration integration}.
\newblock \bibinfo{journal}{\emph{arXiv preprint arXiv:2403.18480}} (\bibinfo{year}{2024}).
\newblock


\bibitem[Wang et~al\mbox{.}(2024c)]%
        {wang2024eager}
\bibfield{author}{\bibinfo{person}{Ye Wang}, \bibinfo{person}{Jiahao Xun}, \bibinfo{person}{Minjie Hong}, \bibinfo{person}{Jieming Zhu}, \bibinfo{person}{Tao Jin}, \bibinfo{person}{Wang Lin}, \bibinfo{person}{Haoyuan Li}, \bibinfo{person}{Linjun Li}, \bibinfo{person}{Yan Xia}, \bibinfo{person}{Zhou Zhao}, {et~al\mbox{.}}} \bibinfo{year}{2024}\natexlab{c}.
\newblock \showarticletitle{EAGER: Two-Stream Generative Recommender with Behavior-Semantic Collaboration}. In \bibinfo{booktitle}{\emph{Proceedings of the 30th ACM SIGKDD Conference on Knowledge Discovery and Data Mining}}. \bibinfo{pages}{3245--3254}.
\newblock


\bibitem[Ying et~al\mbox{.}(2018)]%
        {ying2018graph}
\bibfield{author}{\bibinfo{person}{Rex Ying}, \bibinfo{person}{Ruining He}, \bibinfo{person}{Kaifeng Chen}, \bibinfo{person}{Pong Eksombatchai}, \bibinfo{person}{William~L Hamilton}, {and} \bibinfo{person}{Jure Leskovec}.} \bibinfo{year}{2018}\natexlab{}.
\newblock \showarticletitle{Graph convolutional neural networks for web-scale recommender systems}. In \bibinfo{booktitle}{\emph{Proceedings of the 24th ACM SIGKDD international conference on knowledge discovery \& data mining}}. \bibinfo{pages}{974--983}.
\newblock


\bibitem[Yu et~al\mbox{.}(2023)]%
        {yu2023language}
\bibfield{author}{\bibinfo{person}{Lijun Yu}, \bibinfo{person}{Jos{\'e} Lezama}, \bibinfo{person}{Nitesh~B Gundavarapu}, \bibinfo{person}{Luca Versari}, \bibinfo{person}{Kihyuk Sohn}, \bibinfo{person}{David Minnen}, \bibinfo{person}{Yong Cheng}, \bibinfo{person}{Vighnesh Birodkar}, \bibinfo{person}{Agrim Gupta}, \bibinfo{person}{Xiuye Gu}, {et~al\mbox{.}}} \bibinfo{year}{2023}\natexlab{}.
\newblock \showarticletitle{Language Model Beats Diffusion--Tokenizer is Key to Visual Generation}.
\newblock \bibinfo{journal}{\emph{arXiv preprint arXiv:2310.05737}} (\bibinfo{year}{2023}).
\newblock


\bibitem[Yuan et~al\mbox{.}(2020)]%
        {yuan2020parameter}
\bibfield{author}{\bibinfo{person}{Fajie Yuan}, \bibinfo{person}{Xiangnan He}, \bibinfo{person}{Alexandros Karatzoglou}, {and} \bibinfo{person}{Liguang Zhang}.} \bibinfo{year}{2020}\natexlab{}.
\newblock \showarticletitle{Parameter-efficient transfer from sequential behaviors for user modeling and recommendation}. In \bibinfo{booktitle}{\emph{Proceedings of the 43rd International ACM SIGIR conference on research and development in Information Retrieval}}. \bibinfo{pages}{1469--1478}.
\newblock


\bibitem[Yuan et~al\mbox{.}(2021)]%
        {yuan2021one}
\bibfield{author}{\bibinfo{person}{Fajie Yuan}, \bibinfo{person}{Guoxiao Zhang}, \bibinfo{person}{Alexandros Karatzoglou}, \bibinfo{person}{Joemon Jose}, \bibinfo{person}{Beibei Kong}, {and} \bibinfo{person}{Yudong Li}.} \bibinfo{year}{2021}\natexlab{}.
\newblock \showarticletitle{One person, one model, one world: Learning continual user representation without forgetting}. In \bibinfo{booktitle}{\emph{Proceedings of the 44th International ACM SIGIR Conference on Research and Development in Information Retrieval}}. \bibinfo{pages}{696--705}.
\newblock


\bibitem[Zeng et~al\mbox{.}(2024)]%
        {zeng2024scalable}
\bibfield{author}{\bibinfo{person}{Hansi Zeng}, \bibinfo{person}{Chen Luo}, \bibinfo{person}{Bowen Jin}, \bibinfo{person}{Sheikh~Muhammad Sarwar}, \bibinfo{person}{Tianxin Wei}, {and} \bibinfo{person}{Hamed Zamani}.} \bibinfo{year}{2024}\natexlab{}.
\newblock \showarticletitle{Scalable and effective generative information retrieval}. In \bibinfo{booktitle}{\emph{Proceedings of the ACM on Web Conference 2024}}. \bibinfo{pages}{1441--1452}.
\newblock


\bibitem[Zhai et~al\mbox{.}(2024)]%
        {zhai2024actions}
\bibfield{author}{\bibinfo{person}{Jiaqi Zhai}, \bibinfo{person}{Lucy Liao}, \bibinfo{person}{Xing Liu}, \bibinfo{person}{Yueming Wang}, \bibinfo{person}{Rui Li}, \bibinfo{person}{Xuan Cao}, \bibinfo{person}{Leon Gao}, \bibinfo{person}{Zhaojie Gong}, \bibinfo{person}{Fangda Gu}, \bibinfo{person}{Michael He}, {et~al\mbox{.}}} \bibinfo{year}{2024}\natexlab{}.
\newblock \showarticletitle{Actions speak louder than words: Trillion-parameter sequential transducers for generative recommendations}.
\newblock \bibinfo{journal}{\emph{arXiv preprint arXiv:2402.17152}} (\bibinfo{year}{2024}).
\newblock


\bibitem[Zhang et~al\mbox{.}(2019)]%
        {zhang2019feature}
\bibfield{author}{\bibinfo{person}{Tingting Zhang}, \bibinfo{person}{Pengpeng Zhao}, \bibinfo{person}{Yanchi Liu}, \bibinfo{person}{Victor~S Sheng}, \bibinfo{person}{Jiajie Xu}, \bibinfo{person}{Deqing Wang}, \bibinfo{person}{Guanfeng Liu}, \bibinfo{person}{Xiaofang Zhou}, {et~al\mbox{.}}} \bibinfo{year}{2019}\natexlab{}.
\newblock \showarticletitle{Feature-level deeper self-attention network for sequential recommendation.}. In \bibinfo{booktitle}{\emph{IJCAI}}. \bibinfo{pages}{4320--4326}.
\newblock


\bibitem[Zhao et~al\mbox{.}(2019)]%
        {zhao2019recommending}
\bibfield{author}{\bibinfo{person}{Zhe Zhao}, \bibinfo{person}{Lichan Hong}, \bibinfo{person}{Li Wei}, \bibinfo{person}{Jilin Chen}, \bibinfo{person}{Aniruddh Nath}, \bibinfo{person}{Shawn Andrews}, \bibinfo{person}{Aditee Kumthekar}, \bibinfo{person}{Maheswaran Sathiamoorthy}, \bibinfo{person}{Xinyang Yi}, {and} \bibinfo{person}{Ed Chi}.} \bibinfo{year}{2019}\natexlab{}.
\newblock \showarticletitle{Recommending what video to watch next: a multitask ranking system}. In \bibinfo{booktitle}{\emph{Proceedings of the 13th ACM conference on recommender systems}}. \bibinfo{pages}{43--51}.
\newblock


\bibitem[Zheng et~al\mbox{.}(2024)]%
        {zheng2024adapting}
\bibfield{author}{\bibinfo{person}{Bowen Zheng}, \bibinfo{person}{Yupeng Hou}, \bibinfo{person}{Hongyu Lu}, \bibinfo{person}{Yu Chen}, \bibinfo{person}{Wayne~Xin Zhao}, \bibinfo{person}{Ming Chen}, {and} \bibinfo{person}{Ji-Rong Wen}.} \bibinfo{year}{2024}\natexlab{}.
\newblock \showarticletitle{Adapting large language models by integrating collaborative semantics for recommendation}. In \bibinfo{booktitle}{\emph{2024 IEEE 40th International Conference on Data Engineering (ICDE)}}. IEEE, \bibinfo{pages}{1435--1448}.
\newblock


\bibitem[Zhou et~al\mbox{.}(2018)]%
        {zhou2018deep}
\bibfield{author}{\bibinfo{person}{Guorui Zhou}, \bibinfo{person}{Xiaoqiang Zhu}, \bibinfo{person}{Chenru Song}, \bibinfo{person}{Ying Fan}, \bibinfo{person}{Han Zhu}, \bibinfo{person}{Xiao Ma}, \bibinfo{person}{Yanghui Yan}, \bibinfo{person}{Junqi Jin}, \bibinfo{person}{Han Li}, {and} \bibinfo{person}{Kun Gai}.} \bibinfo{year}{2018}\natexlab{}.
\newblock \showarticletitle{Deep interest network for click-through rate prediction}. In \bibinfo{booktitle}{\emph{Proceedings of the 24th ACM SIGKDD international conference on knowledge discovery \& data mining}}. \bibinfo{pages}{1059--1068}.
\newblock


\bibitem[Zhou et~al\mbox{.}(2020)]%
        {zhou2020s3}
\bibfield{author}{\bibinfo{person}{Kun Zhou}, \bibinfo{person}{Hui Wang}, \bibinfo{person}{Wayne~Xin Zhao}, \bibinfo{person}{Yutao Zhu}, \bibinfo{person}{Sirui Wang}, \bibinfo{person}{Fuzheng Zhang}, \bibinfo{person}{Zhongyuan Wang}, {and} \bibinfo{person}{Ji-Rong Wen}.} \bibinfo{year}{2020}\natexlab{}.
\newblock \showarticletitle{S3-rec: Self-supervised learning for sequential recommendation with mutual information maximization}. In \bibinfo{booktitle}{\emph{Proceedings of the 29th ACM international conference on information \& knowledge management}}. \bibinfo{pages}{1893--1902}.
\newblock


\bibitem[Zhu et~al\mbox{.}(2019)]%
        {zhu2019joint}
\bibfield{author}{\bibinfo{person}{Han Zhu}, \bibinfo{person}{Daqing Chang}, \bibinfo{person}{Ziru Xu}, \bibinfo{person}{Pengye Zhang}, \bibinfo{person}{Xiang Li}, \bibinfo{person}{Jie He}, \bibinfo{person}{Han Li}, \bibinfo{person}{Jian Xu}, {and} \bibinfo{person}{Kun Gai}.} \bibinfo{year}{2019}\natexlab{}.
\newblock \showarticletitle{Joint optimization of tree-based index and deep model for recommender systems}.
\newblock \bibinfo{journal}{\emph{Advances in Neural Information Processing Systems}}  \bibinfo{volume}{32} (\bibinfo{year}{2019}).
\newblock


\bibitem[Zhu et~al\mbox{.}(2018)]%
        {zhu2018learning}
\bibfield{author}{\bibinfo{person}{Han Zhu}, \bibinfo{person}{Xiang Li}, \bibinfo{person}{Pengye Zhang}, \bibinfo{person}{Guozheng Li}, \bibinfo{person}{Jie He}, \bibinfo{person}{Han Li}, {and} \bibinfo{person}{Kun Gai}.} \bibinfo{year}{2018}\natexlab{}.
\newblock \showarticletitle{Learning tree-based deep model for recommender systems}. In \bibinfo{booktitle}{\emph{Proceedings of the 24th ACM SIGKDD international conference on knowledge discovery \& data mining}}. \bibinfo{pages}{1079--1088}.
\newblock


\bibitem[Zhu et~al\mbox{.}(2024)]%
        {zhu2024cost}
\bibfield{author}{\bibinfo{person}{Jieming Zhu}, \bibinfo{person}{Mengqun Jin}, \bibinfo{person}{Qijiong Liu}, \bibinfo{person}{Zexuan Qiu}, \bibinfo{person}{Zhenhua Dong}, {and} \bibinfo{person}{Xiu Li}.} \bibinfo{year}{2024}\natexlab{}.
\newblock \showarticletitle{CoST: Contrastive Quantization based Semantic Tokenization for Generative Recommendation}. In \bibinfo{booktitle}{\emph{Proceedings of the 18th ACM Conference on Recommender Systems}}. \bibinfo{pages}{969--974}.
\newblock


\end{thebibliography}
\end{document}